\documentclass[twocolumn]{revtex4-1}

\usepackage{graphicx}
\usepackage{dcolumn}
\usepackage{bm}

\usepackage{graphicx}
\usepackage{pstricks,pst-node,pst-text,pst-3d}
\usepackage{amssymb}
\usepackage{xspace}
\usepackage{amsmath}
\usepackage{units}
\usepackage{picinpar}
\usepackage{subfigure}
\usepackage{url,multirow}

\begin{document}

\preprint{XXXXXXXX}

\title{Hadronic Multiparticle Production at
  Ultra-High Energies and Extensive Air Showers}
\author{Ralf~Ulrich\footnote{Corresponding author,
    rulrich@phys.psu.edu}$^{\rm a}$} \author{Ralph~Engel$^{\rm b}$}
\author{Michael~Unger$^{\rm b}$} \affiliation{%
  $^{\rm a}$ The Pennsylvania State University, Center for Particle
  Astrophysics, 104 Davey Lab, University Park, PA 16802, USA}
\affiliation{%
  $^{\rm b}$ Karlsruhe Institute of Technology (KIT), Institut f\"ur
  Kernphysik, P.O. Box 3640, 76021 Karlsruhe, Germany } \date{\today}

%
\begin{abstract}  
  Studies of the nature of cosmic ray particles at the highest energies
  are based on the measurement of extensive air showers.
  Most cosmic ray properties can therefore 
  only be obtained from the interpretation of air
  shower data and are thus depending on predictions of hadronic
  interaction models at ultra-high energies.  We discuss different
  scenarios of model extrapolations from accelerator data to
  air shower energies and investigate their impact on the
  corresponding air shower predictions.
  To explore the effect of different extrapolations by hadronic interaction
  models we developed an ad hoc model.  This ad hoc model is based on
  the modification of the output of standard hadronic interaction event generators within
  the air shower simulation process and allows us to study the impact 
  of changing interaction features on the air shower development.
  In a systematic study we
  demonstrate the resulting changes of important air shower
  observables and also discuss them in terms of the predictions of the
  Heitler model of air shower cascades.
  It is found that the results of our ad hoc modifications are, to
  a large extend, independent of the choice of the underlying hadronic
  interaction model.

\end{abstract}

\maketitle


\section{Introduction}
The nature of the highest energy cosmic ray particles is still an open
question.  Even with recent high exposure experiments delivering large
quantities of data at ultra-high
energies~\cite{Abraham:2004dt,Kawai:2008zza,Abbasi:2004nz}, the
fundamental problem remains to interpret the collected air shower
data. This is mainly due to difficulties and limitations in
modelling of hadronic interactions in air shower cascades. Whereas we
can predict reliably electromagnetic interactions with QED, it is still
not possible to calculate hadronic multiparticle production from first
principles. Moreover, most of the interaction energies and phase space
regions of relevance to the development of air showers are not directly
accessible in accelerator experiments.  Thus, phenomenological models
of hadronic interactions have to be used to extrapolate the available
accelerator measurements to these unexplored phase space regions. 

Due to the lack of a theoretical framework that allows one to make
quantitative predictions, the systematic uncertainties of these
extrapolations are very difficult to estimate.  This is a well known
problem for, e.g.\ the determination of the systematic
uncertainty of the primary mass composition derived from air shower
measurements~\cite{Dawson:1998kk,Ave:2002gc,Dova:2003an,Abbasi:2004nz,%
  Abraham:2009ds,Abraham:2010yv,Antoni:2005wq}. Similarly, the
reconstruction of the energy of the primary particle from detector
data calibrated by Monte Carlo simulations is subject to largely unknown model
uncertainties~\cite{Antoni:2005wq,Nagano:1992jz,Takeda:1998ps,%
  Ave:2001hq, Aglietta:2004np, Amenomori:2008jb}.  

For estimating systematic uncertainties of air shower predictions, one
first needs to know the uncertainties of the model predictions for
multiparticle production, and secondly how these uncertainties propagate
to predictions for different air shower observables. In this work we
will address the latter by studying the relation of different
extrapolations of hadronic particle production to air shower
predictions. 

We will briefly discuss the uncertainties associated with the modelling
of hadronic interactions and argue that the differences of existing
models do not cover the full range of the expected uncertainty at
energies beyond the reach of colliders. To nevertheless explore a wide
range of different extrapolations and to calculate the corresponding
air shower observables, we will introduce an ad hoc modification of
existing models. The shower observables under consideration are
the depth of shower maximum and the number of muons at ground level--as the
model-dependence of the interpretation of air shower data is mainly
related to these two parameters~\cite{Heck08CORSIKASchool}. All other
characteristics of air showers are closely related to those two
features due to universality of the electromagnetic shower component
(see \cite{Rossi:1941zz,Giller:2005qz,Nerling:2006yt,Schmidt:2007vq,Lipari:2008td,Lafebre:2009en} and 
Refs.\ therein). The
results will also be compared to the expectations from the Heitler-Matthews
cascade model~\cite{Matthews:2005sd}.

\begin{figure*}[t!] 
  \centering
  \includegraphics[width=.45\linewidth]{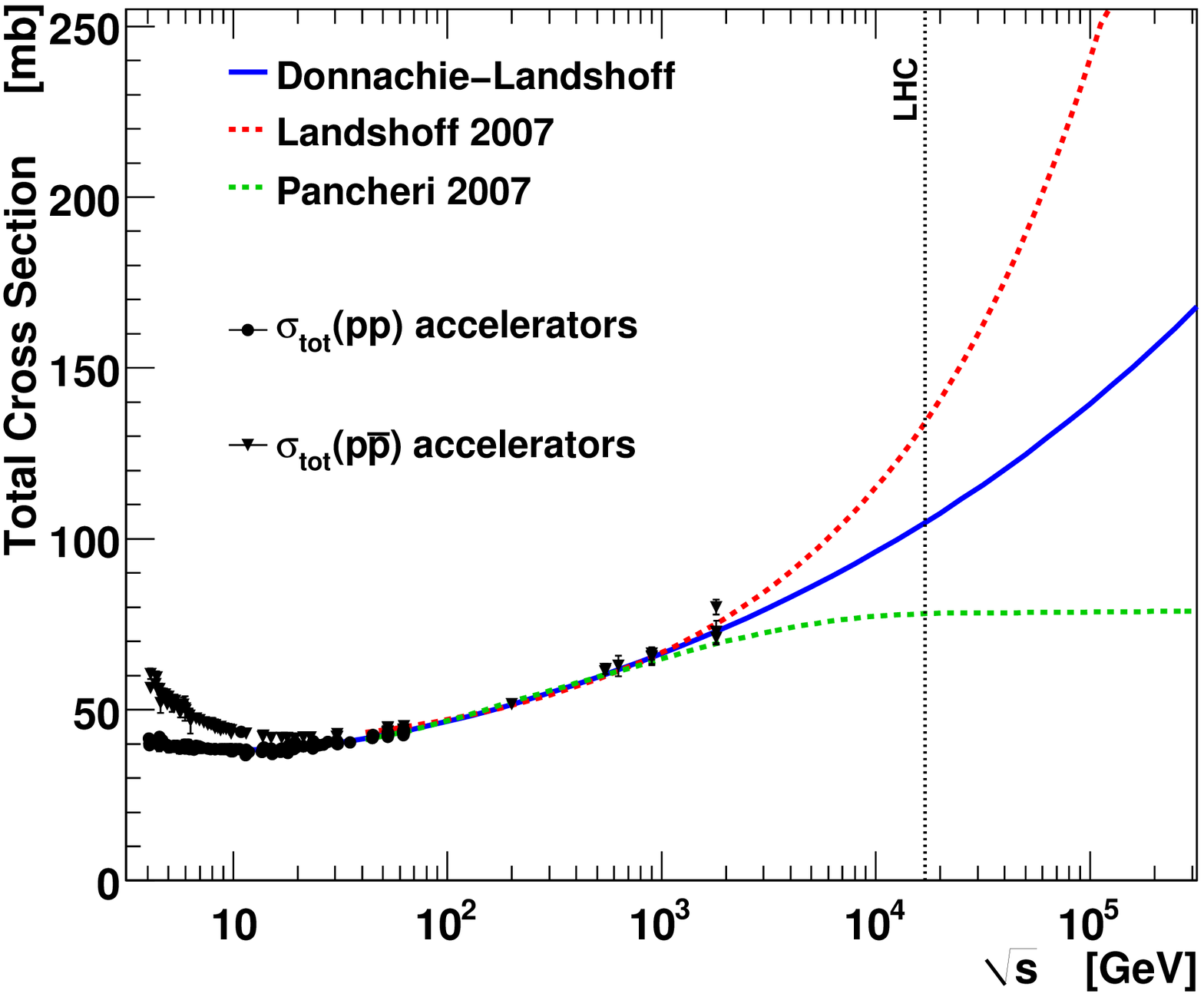}~
  \hfill~
  \includegraphics[width=.45\linewidth]{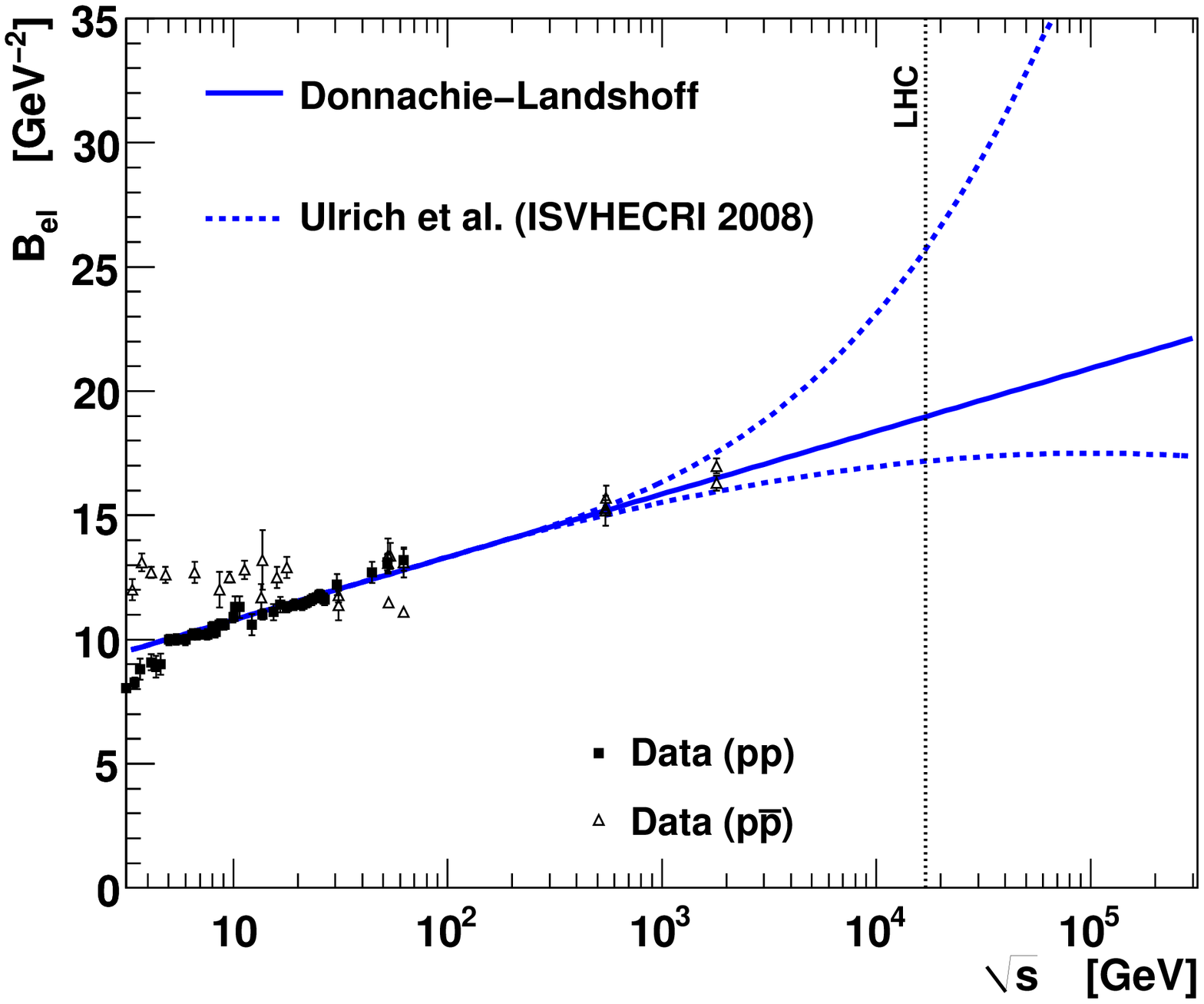}
  \caption{Compilation of accelerator data of $\sigma^{\rm pp}_{\rm
      tot}$ and $B_{\rm el}$~\protect\cite{Engel:2000pb}. The central
    line denotes the conventional extrapolation of these data to high
    energy. The upper and lower lines indicate a set of possible
    extreme extrapolations.  In the left plot the conventional model
    is the soft pomeron parametrization by Donnachie and
    Landshoff~\protect\cite{Donnachie:1992ny}, while the lower curve
    is by Pancheri et al.~\protect\cite{Pancheri:2007rv} and the upper
    one is the two-pomeron model of
    Landshoff~\protect\cite{Landshoff:2007uk,Landshoff:2009wt}. The
    different scenarios in the right plot are
    from~\cite{Ulrich:2009yq}.}
  \label{fig:sigmaExtrapolations}
\end{figure*} 

\section{Uncertainties in Modelling Hadronic Interactions}
\label{sec:uncert}
The development of extensive air showers (EAS) is very sensitive to
the characteristics of hadronic interactions at ultra-high energies.
Different hadronic interaction event generators are available for the
simulation of EAS. 

Currently it is unclear whether the existing differences in model
predictions for hadronic multiparticle production can be used to
estimate the level of theoretical uncertainties related to our limited
understanding of hadronic interactions in air showers. It is possible
that existing model-differences are
\begin{enumerate}
 \item[(A)] \emph{larger} than the actual systematic
  uncertainties. For example, progress in understanding hadronic particle
  production and new data from accelerators allow us to update
  interaction models to obtain a more realistic description of
  particle production. Not all models are updated regularly and the
  quality of data description differs between the models.
 \item[(B)] \emph{smaller} than the actual systematic
  uncertainties. The existing models are not covering
  the full phase space of possible interaction scenarios and
  parameters. Moreover, new physics processes at higher energies,
  which are unknown now and thus missing in current modelling
  approaches, could change extrapolations drastically.
\end{enumerate}

Frequently used models for the high-energy range are
\textsc{QGSJet~II}~\cite{Ostapchenko:2004ss,Ostapchenko:2006vr},
\textsc{Epos}~\cite{Werner:2007vd},  and the somewhat older 
\textsc{QGSJet~01}~\cite{Kalmykov:1993qe,Kalmykov:1989br} and
\textsc{Sibyll~2.1}~\cite{Ahn:2009wx}. These models are available in
the air shower simulation package CONEX~\cite{Bergmann:2006yz} that
will be used for calculating the shower observables. Other models for
hadronic interactions include
\textsc{neXus}~\cite{Drescher:2000ha,Hladik:2001zy},
\textsc{HDPM}~\cite{Capdevielle:1989ht},
\textsc{DPMJET}~\cite{Ranft:1994fd}, and
\textsc{VENUS}~\cite{Werner:1993uh}. These models are older or more
limited in the scope of application and not considered here. 

Despite the different level of sophistication, the predictions by
\textsc{Sibyll~2.1} and \textsc{QGSJet~01} are not objectively worse
than those by \textsc{QGSJetII} and \textsc{Epos}, as many model
aspects are assumptions that cannot be justified by underlying
fundamental theoretical
constraints.  Over the years model predictions and extrapolations have
become more alike even though there is no theory for calculating e.g.\
cross sections from first principles~\cite{Heck01a}. One has to be
careful and should not consider this increasing similarity of model
predictions as real convergence and significant decrease of the
uncertainties. None of the models is able to consistently describe
cosmic ray data (e.g.~\cite{Antoni:2001pw,Antoni:2005wq,AbuZayyad:1999xa,Abraham:2009ds}). In the
energy range up to about $\unit[10^{15}]{eV}$ where various
measurements on multiparticle production are available good tuning to
many different data sets should indeed lead to a convergence of the
model predictions. However, at energies beyond that of collider
experiments, the extrapolations can only be guided by theoretical end
phenomenological assumptions.

\begin{figure}[b] 
  \centering
  \includegraphics[width=\linewidth]{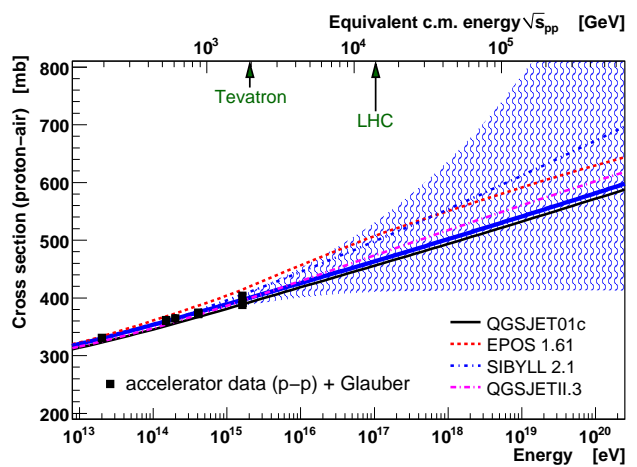}
  \caption{Uncertainty of the extrapolation of the proton-air cross
    section for particle production due to different models of the proton-proton cross
    section as calculated with the Glauber
    framework~\cite{Ulrich:2009yq}.}
  \label{fig:sigmaExtrapolationsEAS}
\end{figure} 

In Fig.~\ref{fig:sigmaExtrapolations} accelerator data on
the total proton-proton cross section $\sigma_{\rm tot}^{\rm pp}$ and
the elastic slope parameter $B_{\rm el}$, defined by ${\rm
  d}\sigma_{\rm el}/{\rm d}t|_{t=0}\propto\exp(-|t|B_{\rm el})$, are
shown together with different models that extrapolate these data to
ultra-high energies. Converting these model extrapolations within the Glauber
framework~\cite{Glauber:1955qq,Glauber:1970jm} to
proton-air cross sections for particle production as needed for air
shower simulation leads to a wide range
of predictions at ultra-high energy~\cite{Ulrich:2009yq}.  This range
is shown in Fig.~\ref{fig:sigmaExtrapolationsEAS} as the shaded area. The
extrapolations of aforementioned interaction models are also shown and it is obvious
that they do not exhaust the possible range of uncertainties.  In
addition it can be seen that a measurement of the total cross section
at LHC energy with an uncertainty smaller than a few percent has the
potential to significantly reduce the uncertainties of the model
extrapolations at cosmic ray energies. Furthermore, recent activities of
model development of the \textsc{Epos} event generator clearly
demonstrated that there exists a large freedom to change
predictions at ultra-high energies~\cite{Pierog:2006qv,Pierog:2007x1}.

All this is demonstrating that the current
situation is most likely such that one would underestimate the systematic
uncertainties of model extrapolations at energies beyond the reach
of accelerator experiments if one would just consider the
extrapolation of existing event generators. This is supported by the
fact that existing interaction models are not able to consistently describe
cosmic ray observations. Therefore it is also not surprising that
first LHC
data~\cite{Aamodt:2010dx,Aamodt:2010ft,Aamodt:2010pp,Aamodt:2009dt,%
Aad:2010rd,%
Khachatryan:2010xs,Khachatryan:2010us,Khachatryan:2010pv}
indicate some model deficiencies~\cite{dEnterria:2010x1}.

%
%

In the following we will consider different extrapolations of
multiparticle production to energies higher than
$\unit[10^{15}]{eV}$, corresponding to a c.m.s.\ energy of about $\unit[2]{TeV}$
for proton-proton collisions. We will concentrate on general features of hadronic particle
production that are most directly linked to air shower predictions.

%
\section{Predictions of the Cascade Model for Air Showers}
Simple cascade models, often referred to as Heitler
models~\footnote{For a historical account of the development of this
  model see~\protect\cite{Linsley:1978zg}.}, are
providing some insight into how air shower observables are related to
interaction physics on a microscopic
level~\cite{RevModPhys.21.113,Matthews:2005sd,Pierog:2006qu}.  The
descriptive strength of Heitler models--despite their extreme
simplicity--is remarkable. We will discuss the ability of such
models to link the physics of interactions to air shower observables,
but also point out the limitations of these models.

%
\subsection{Electromagnetic Heitler Model}
\begin{figure}[hbt]
  \centering
  \includegraphics[width=.6\linewidth]{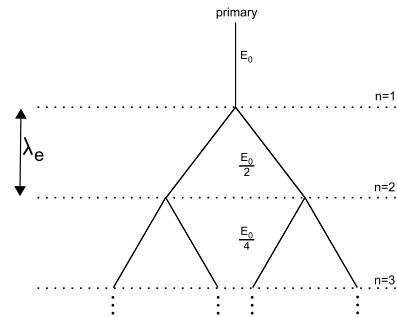}~  
  \caption{Electromagnetic Heitler model.}
  \label{fig:Heitler}
\end{figure}

In the electromagnetic Heitler approximation only one particle type is
considered and substitutes $\gamma$, e$^+$ and e$^-$. It is assumed
that an e.m.\ particle with energy $E$ interacts after one splitting
length $\lambda_{\rm e}=\ln2\,X_0$, where $X_0\sim\unit[37]{g/cm^2}$
is the e.m.\ radiation length, producing two secondaries with energy
$E/2$, see Fig.~\ref{fig:Heitler}.
The number of particles after each
splitting length increases by a factor of two, and thus the number of
particles at generation $n$ is
\begin{equation}
  N_n = 2^n
\end{equation}
and their energy is
\begin{equation}
  E_n = E_0 / N_n  \;.
\end{equation}
Defining the critical energy $E_{\rm c}^{\rm e.m.}$ as the energy
below which energy loss processes dominate over particle production
($\sim$ $85\,$MeV in air), one can make the assumption that the shower
maximum is reached when the energy of secondary particles reaches 
$E_{\rm c}^{\rm e.m.}$. Thus, two main shower observables are given by
\begin{eqnarray}
  N_{\rm max} &=& E_0 / E_{\rm c}^{\rm e.m.} \sim E_0 \qquad\quad\text{and}\nonumber\\
 X_{\rm max}&=&\lambda_{\rm e} n_{\rm c} \sim \lambda_{\rm e}\ln(E_0)
 \;,
\end{eqnarray}
with $n_{\rm c}=(\ln2)^{-1}\,\ln\left(E_0/E_{\rm c}^{\rm
    e.m.}\right)$.  Even this very simplistic model reproduces two
important features of air showers: the number of particles at shower
maximum $N_{\rm max}$ is proportional to $E_0$ and the depth of shower
maximum $X_{\rm max}$ depends logarithmically on the primary energy $E_0$.

\subsection{Hadronic Extension of the Heitler Model}
\label{sec:hadronicHeitler}

\begin{figure}[hbt]
  \centering
  \includegraphics[width=.6\linewidth]{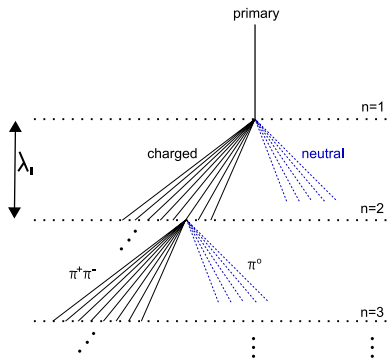}
  \caption{Hadronic Heitler-Matthews model: pion cascade in air.}
  \label{fig:HeitlerExt}
\end{figure}

The Heitler model can be extended to hadronic particles by considering
a cascade of pions interacting in air~\cite{Matthews:2005sd}. It is assumed that a charged
pion ($\pi^+$ or $\pi^-$) of energy $E$ interacts after one
interaction length $\lambda_{\rm I}$ ($\unit[\approx120]{g/cm^2}$ for
pions of $\unit[10-1000]{GeV}$~\cite{Gaisser:1990vg}) and produces $r\,
n_{\rm mult}$ charged pions and $c\,n_{\rm mult}$ neutral pions, see
Fig.~\ref{fig:HeitlerExt}. Here $c=1-r$ is the pion charge-ratio
and is in the Heitler framework typically defined to be $1/3$. 
Neutral pions decay instantaneously into $2\gamma$ and are thus lost
from the hadronic cascade, while transferring  energy into the
electromagnetic cascade. 
The number of charged pions in the hadronic cascade at generation $n$
is
\begin{equation}
  N^{\rm ch}_n = (r\,n_{\rm mult})^n
\end{equation}
and the energy per pion 
\begin{equation}
  E^\pi_n = \frac{E_0}{(n_{\rm mult})^n} \;. 
\end{equation}
It is assumed that charged pions decay into muons as soon as
their energy drops below $E_{\rm dec}$. This is the characteristic energy where the
interaction length becomes larger than the decay length of pions in
air, and is thus also depending on the atmospheric density
profile. For example high up in the atmosphere at a depth of
$\unit[\sim100]{g/cm^2}$ this energy is around $E_{\rm
  dec}\sim55\,$GeV and decreases to $E_{\rm dec}\sim18\,$GeV at sea
level. In realistic air shower scenarios the energy $E^\pi_n=E_{\rm dec}$ 
is typically reached within the range of
$\unit[20-30]{GeV}$~\cite{Gaisser:1990vg,Matthews:2005sd,Alvarez-Muniz:2002ne}, 
which is then adopted to
be the \emph{critical energy} of pions $E_{\rm c}^\pi$, in analogy to
the critical energy of electrons $E_{\rm c}^{\rm e.m.}$. The number of
generations required to reach $E^\pi_n=E_{\rm c}^\pi$ is 
\begin{equation}
  n_{\rm c} = \frac{\ln(E_0/E_{\rm c}^\pi)}{\ln n_{\rm mult}}\;,
\end{equation}
and is for air showers in the range of $4$ to $7$ \cite{Meurer:2005dt} . This yields a muon
number from decaying charged pions of
\begin{eqnarray}
  N_\mu &=& N_{n_{\rm c}}^{\rm ch}= (r\,n_{\rm mult})^{n_{\rm c}} = \left(\frac{E_0}{E_{\rm c}^\pi}\right)^\beta \quad\qquad\text{with} \nonumber\\
  \beta &=& \frac{\ln r\,n_{\rm mult}}{\ln n_{\rm mult}}=\frac{\ln r}{\ln n_{\rm mult}}+1 \;, 
\end{eqnarray}
which depends on the
secondary particle multiplicity and the pion charge ratio of hadronic
interactions. Moreover, the electron number at shower maximum can be
estimated from
\begin{equation}
  \label{eqn:NmaxeHeitler}
  N_{\rm max,\,e}=\frac{E_{\rm e.m.}}{E_{\rm c}^{\rm e.m.}} = \frac{E_0}{E_{\rm c}^{\rm e.m.}} - N_\mu \frac{E_{\rm c}^{\pi}}{E_{\rm c}^{\rm e.m.}} \;,
\end{equation}
and is thus depending on the hadronic secondary multiplicity and pion
charge ratio via the muon number.  To relate the electron number at
the shower maximum to the electron number at observation level,
$N_{\rm e}$, it is necessary to take the strong attenuation of the
electromagnetic shower component after the shower maximum into
account. In the limit where $X_{\rm obs}\gg X_{\rm max}$ the air
shower size is in very good approximation exponentially attenuated
with the scale length $\Lambda\sim\unit[65]{g/cm^2}$~\footnote{It is a feature
of air shower universality~(see, for example,
\cite{Rossi:1941zz,Giller:2005qz,Nerling:2006yt,Schmidt:2007vq,Lipari:2008td,Lafebre:2009en}) that $\Lambda$ does
not depend on the primary particle or hadronic interaction model.}.
We can thus use
\begin{equation*}
  N_{\rm e} \sim N_{\rm e,\,max}\,e^{-(X_{\rm obs}-X_{\rm max})/\Lambda}\quad \text{for}\quad X_{\rm obs}\gg X_{\rm max}\,,
\end{equation*}
which leads to
\begin{equation}
  \label{eqn:NeHeitler}
  \ln N_{\rm e}\sim \ln N_{\rm e,\,max} - (X_{\rm obs}-X_{\rm max})/\Lambda\,.
\end{equation}
It is important to note that $\ln N_{\rm e}$ depends only
logarithmically on $N_{\rm e,\,max}$ but linearly from $X_{\rm
  max}$. It is shown in Section~\ref{sec:results} that indeed the impact
of the depth of $X_{\rm max}$ on the electron number at ground level
is dominating over a change following from
Eq.~(\ref{eqn:NmaxeHeitler}). It is thus the strong longitudinal shower evolution 
that is responsible for the inability of the Heitler model to directly infer 
any dependence of $N_{\rm e}$ on hadronic interaction characteristics.

Within the hadronic Heitler framework it is not possible to follow the parallel
development of the hadronic and electromagnetic cascades. To
estimate the position of the shower maximum 
one is forced to consider the electromagnetic contribution from
the first hadronic interaction only, which consists of $2\,c\,n_{\rm
  mult}$ e.m.\ subshowers each of the energy $E_0/n_{\rm
  mult}$. The shower maximum is then
\begin{equation}
  \label{eqn:hadXmax}
  X_{\rm max} \sim \lambda_{\rm I} + \lambda_{\rm e} \ln \frac{E_0}{n_{\rm mult}\,E_{\rm c}^{\rm e.m.}} \;,
\end{equation}
which is proportional to the hadronic interaction length and depends
logarithmically on the multiplicity, but not on the charge ratio.
This expression does not include the contribution to the electromagnetic
cascade from the subsequent hadronic cascading process.
However, the inclusion of higher
hadronic generations does not change the structure of
Eq.~(\ref{eqn:hadXmax}), only the coefficients change
(e.g.~\cite{AlvarezMuniz:2002ne}).

\subsection{Inelasticity in the Heitler Model}

In the Heitler model only equal energy particles of one type
are considered. This also excludes any account of leading particle
effects or other secondary particle distribution effects. To some
limited extend it was accomplished to incorporate the inelasticity of
interactions in a Heitler-type cascade by
Matthews~\cite{Matthews:2005sd}. The inelasticity
\begin{equation}
\kappa_{\rm  inel}=1-E_{\rm leading}/E_0 = 1- \kappa_{\rm el}
\end{equation}
is the fraction of the primary energy that
is not carried away by the most energetic secondary particle, often
referred to as leading particle.
This energy fraction is available for the production of new secondary
particles, mainly pions and kaons. The elasticity
$\kappa_{\rm el}=1-\kappa_{\rm inel}$ is the fraction of the primary energy
that is carried by the leading particle.

The main difference compared to the standard Heitler model is that
after each interaction secondaries with two energy levels are
generated: $n_{\rm mult}$ particles with energy of $\kappa_{\rm
  inel}\,E_0/n_{\rm mult}$ (of which $c\,n_{\rm mult}$ are neutral
pions) and one particle with energy of $\kappa_{\rm el}\,E_0$. So with
an increasing number of hadronic generations the particles are
distributed in more an more distinct energy levels. It turns out that in the
generation $n$ there are in fact $n+1$ distinct groups of particles of
identical energy. This situation is too complex to be handled in a
compact analytic way.
According to Matthews almost all secondary particles populate the one
or two lowest energy levels per generation. Thus it can be justified
to proceed as in the normal hadronic Heitler model and approximate the
terminal hadronic generation $n_{\rm c}$ as where the \emph{average}
energy per pion drops below the critical energy, and thus
\begin{equation}
  n_{\rm c} = \frac{\ln(E_0/E_{\rm c}^\pi)}{\ln\left((1+r\,n_{\rm mult})/(1-c\,\kappa_{\rm inel})\right)}\;,
\end{equation}
where the total number of pions in generation $n$ is
\begin{equation}
  N_n^\pi = (1 + r\,n_{\rm mult})^n
\end{equation}
and the average energy of these pions is
\begin{equation}
  E_n^\pi = E_0 (1 - c\,\kappa_{\rm inel})^n / N_n^\pi \;.
\end{equation}
The muon number is then
\begin{eqnarray}
  \label{eqn:NmuHeitler}
  N_\mu &=& \left(\frac{E_0}{E_{\rm c}^\pi}\right)^\beta \qquad \text{with}\nonumber\\
  \beta &=& \frac{\ln(1+r\,n_{\rm mult})}{\ln\left((1+r\,n_{\rm mult})/(1-c\,\kappa_{\rm inel})\right)}\nonumber\\ &=& \left(1-\frac{\ln(1-c\,\kappa_{\rm inel})}{\ln(1+r\,n_{\rm mult})} \right)^{-1}\;.
\end{eqnarray}

Since the inelasticity of hadronic interactions in air showers is of
the order of $0.6$ the approximation that the electromagnetic output
of the first interaction is dominating the depth of the shower maximum
is in fact not justified. Leading particles carry a large
fraction of the total energy deeper into the atmosphere and
contribute significantly to the electromagnetic cascade by finally generating
e.m.\ sub-showers at larger depths. This superposition of electromagnetic subshowers
starting at different hadronic generations does not fit into the
analytical frame of the Heitler model, and thus no relation between
inelasticity and $X_{\rm max}$ can be given. In fact, a simple
extension of Eq.~(\ref{eqn:hadXmax})
suggests that the value of $X_{\rm max}$ would \emph{decrease} proportional
to $\lambda_{\rm e}\,\ln (1-\kappa_{\rm el})$, while, on the  contrary,
it should \emph{increase} with increasing elasticity because more energy is transferred deeper
into the atmosphere. This is confirmed by our calculations, 
cf.\ Section~\ref{sec:results}. 
%
In all these considerations we have not taken into account the fact
that the interaction cross sections, the elasticity, the secondary
particle multiplicity and many more parameters of hadronic
interactions depend on the energy. It is possible
to improve the Heitler model to include some of these effects but this
goes beyond the scope of this work (see, for example, \cite{Alvarez-Muniz:2002ne}) .
\begin{table}
  \caption{Proportionalities found in the extended Heitler model
    between hadronic interaction features and the average of air
    shower observables. Some of the proportionalities are
    approximations that also depend on the value of other
    parameters.}
  \begin{tabular}{l||c|c|c}
    & $X_{\rm max}$ & $\ln N_{\mu}$ & $\ln N_{\rm e,\,max}$ \\
    \hline
    \hline
    hadr.\ cross section & $1/\sigma_{\rm I}$ & 1 & 1 \\
    multiplicity & $-\ln n_{\rm mult}$ & $\sim-1/\ln n_{\rm mult}$ & $\sim1/\ln n_{\rm mult}$ \\  
    elasticity & N/A & $\sim\kappa_{\rm el}$ & $\sim-\kappa_{\rm el}$ \\
    pion charge ratio & 1 & $\sim-c$ & $\sim c$ \\
  \end{tabular}
  \label{tab:heitler}
\end{table}

In Table~\ref{tab:heitler} the expected relations between features of hadronic
interactions and important air shower observables are summarized. The
results for $X_{\rm max}$ are all obtained from
Eq.~(\ref{eqn:hadXmax}), where the relation $\lambda_{\rm
  I}\propto1/\sigma_{\rm I}$ is used.
The results for the muon numbers are approximations of
\begin{equation}
  \label{eqn:LogNmuHeitler}
  \ln N_\mu=\beta \ln(E_0/E_{\rm c}^{\pi})
\end{equation}
based on Eq.~(\ref{eqn:NmuHeitler}). For comparison, Matthews
quotes for constant $E_0$ that
$\ln N_\mu\propto\beta\approx1-0.14\kappa_{\rm inel}$, which corresponds
to our result.
The change of the electron numbers at the shower maximum
is given by ${\rm d}\ln N_{\rm
  e,\,max}={\rm d}N_{\rm e,\,max}/N_{\rm e,\,max}\propto-{\rm d}N_{\mu}/N_{\rm
  e,\,max}=-N_\mu/N_{\rm e,\,max}\,{\rm d}\ln N_{\mu}\propto-{\rm d}\ln
N_{\mu}$, which is derived from Eq.~(\ref{eqn:NmaxeHeitler}).
We will confront the results listed in this table with our simulations
in Section~\ref{sec:results}.

\subsection{Fluctuations of the Air Shower Development}
\begin{figure}[b]
  \centering
  \includegraphics[width=\linewidth]{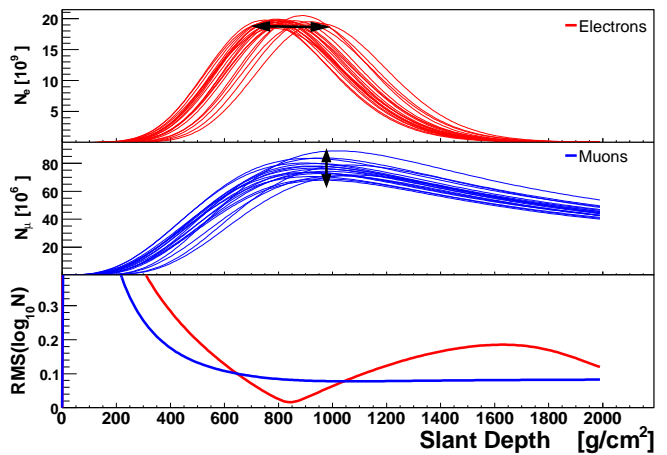}
  \caption{Fluctuation of the longitudinal air shower development for
    20 proton induced EAS at $E_0=\unit[10^{19.5}]{eV}$. }
  \label{fig:protonFluctuations}
\end{figure} 

Not only the mean values of air shower observables are important, but also
the fluctuations of these observables contain valuable
information. They can complement the interpretation of air showers by
exploiting different aspects of the shower development. Compared to
average shower properties, the investigation of shower fluctuations
has the potential to yield more reliable results.

The Heitler model has also some limited capabilities to explain the
fluctuations of air shower cascades. According to
Eq.~(\ref{eqn:hadXmax}) fluctuations in the depth of the shower
maximum can be related to fluctuations of the depth of the first
interaction ${\rm V}[X_1]=\lambda_{\rm I}^2\propto1/\sigma_{\rm I}^2$
and fluctuations of the multiplicity via
\begin{align}
  {\rm V}[X_{\rm max}] &\propto 1/\sigma_{\rm I}^2 + \lambda_{\rm e}^2\,{\rm V}[\ln n_{\rm mult}]\,.
  \label{eq:heitlerVar}
\end{align}
Thus, for small $\sigma_{\rm I}$, the fluctuations in $X_{\rm max}$
are depending on the cross section of the cosmic ray
projectile in the atmosphere, while for very large $\sigma_{\rm I}$ the
fluctuations of the multiplicity are
dominating.

The fluctuations of the logarithm of the muon number can be derived
from Eq.~(\ref{eqn:LogNmuHeitler}). The resulting coefficients are complicated
functions of $c$, $k_{\rm inel}$, $n_{\rm mult}$ and $E_{\rm c}^\pi$,
and it is not helpful to discuss them further.
Fluctuations of the electron number are directly linked to the
fluctuations of the muon number and $X_{\rm max}$ by 
Eqs.~(\ref{eqn:NmaxeHeitler}) and (\ref{eqn:NeHeitler}).
It is demonstrated in Section~\ref{sec:results} that the
correlation between the
fluctuations in the longitudinal shower development, i.e. the depth of the
shower maximum, and the 
fluctuations of the electron number is very strong and mostly
dominates over other effects.

To get an impression on important features of fluctuations of the
longitudinal shower development of electron and muon number profiles,
we perform some more detailed simulations beyond the Heitler model.

The characteristic fluctuations of air showers are demonstrated in
Fig.~\ref{fig:protonFluctuations}, where electron and muon number
profiles are shown for a series of equal energy air showers simulated
with CONEX~\cite{Bergmann:2006yz}. The depth of the electron number
profiles shows large fluctuations while the electron number at the
maximum is almost constant. On the other hand, the depth of the
maximum in the muon profile does not show large fluctuations, but the
maximum number of muons does. Also the electron profile has a very
clear maximum with a fast absorption at larger depth, while the muon
profile only exhibits a very moderate absorption after the
maximum. This is also reflected in the logarithmic RMS of the electron
and muon profiles around their mean profile. The electromagnetic part
of the shower exhibit a very pronounced minimum in its fluctuations at
the depth of the mean shower maximum $\langle X_{\rm max}\rangle$.

\section{Simulation of Air Showers with Modified Interaction Models}
\label{sec:method}

In earlier studies of similar aim it was investigated how the
interpretation of air showers is affected by changing internal
parameters of individual hadronic interaction
models~\cite{Bellandi:1995pj,Engel:2003ac,Ostapchenko:2003sj,Ostapchenko:2006ks,%
  Bleve:icrcLodz} or by comparing the predictions of different interaction
models
\cite{Knapp96a,Knapp:2002vs,Anchordoqui:1998nq,GarciaCanal:2009xq}.
These approaches have the virtue of not, or only marginally, leaving
the allowed parameter space of the original models. However,
underlying assumptions that cannot be justified by fundamental
theoretical principles, on which all models are built on, are often
even more important than just parameter values. Because of this, only
the phase-space in the direct vicinity of the original model can be
explored by parameter variations. It is found that typically the
differences between different models are larger than what can be
obtained from variations of internal model parameters. Furthermore the
available models do not exhaust the full range of current theoretical
uncertainties.

This is why for our study we are not changing internal parameters of
hadronic interaction models, but developed an ad hoc model to explore
the uncertainties beyond the predictions of the available interaction
models.  Our model uses the predictions of existing hadronic event
generators and modifies the output of these models in a suitable way
to probe the phase-space of interaction characteristics more
exhaustively.  These modifications are ad hoc and explicitly
\textit{not} based on an underlying fundamental theory or
phenomenology.  This allows us, firstly, to apply our modification to
any interaction model available and, secondly, to scan the phase-space
of interaction physics very freely and extensively.

It is one of the advantages of our approach that it is not bound to
parameter correlations normally arising from model-internal
mechanisms. Therefore we can study the influence of the modification
of one parameter on the shower evolution almost independently of the
other parameters. For example, the increase of the inelasticity is typically correlated
to an increase of the secondary particle multiplicity. This is not the
case in our way of modifying the model predictions.
Of course, energy and charge conservation imposes
some correlation between different parameters that are present even
in our approach. 

\subsection{Considered Parameters of Hadronic Interactions}

We investigate the impact of the variation of several important features of
hadronic interactions on typical air shower observables. These are
\begin{itemize}

\item the \emph{secondary multiplicity}, $n_{\rm mult}$, which is
  defined as the total number of secondaries produced in a hadronic
  collision,

\item \emph{hadronic particle production cross sections}, $\sigma_{\rm prod}$,
 that determine the interaction length of particles and thus the 
 depth of the first interaction in the atmosphere, but also the
 speed of the development of the hadronic shower core,

\item the \emph{elasticity}, $\kappa_{\rm el}=E_{\rm leading}/E_{\rm tot}$ with
  $E_{\rm leading}$ being the energy in the lab system carried by the
  leading particle after the collision and $E_{\rm tot}$ the energy
  of the projectile in the lab system,

\item and the \emph{pion charge-ratio},
  $c=n_{\pi^0}/(n_{\pi^0}+n_{\pi^+}+n_{\pi^-})$, with
  $n_{\pi^{\rm x}}$ the number of pions of type $\pi^{\rm x}$,
  which determines the fraction of particles going into the
  electromagnetic cascade of the air shower after each interaction,
  assuming that all $\pi^0$ decay basically instantly into $2\gamma$.
\end{itemize}

The air shower observables that we study are the shower maximum,
$X_{\rm max}$, the total number of electrons above \unit[1]{MeV},
$N_{\rm e}$, as well as muons above $\unit[1]{GeV}$, $N_{\mu}$,
arriving at the slant depth of $X_{\rm obs}=\unit[1000]{g/cm^{2}}$,
and the invisible energy, $E_{\rm inv}$, which is not accessible to
calorimetric fluorescence telescope measurements.

\subsection{Extrapolation in Energy}

To investigate the importance of the extrapolation of high energy hadronic
interaction models for the interpretation of EAS observables, we
modified interaction characteristics during EAS simulations. To
achieve this, individual hadronic interaction features are
altered by the energy-dependent factor
\begin{align}
  \label{eqn:modifier}
  f(E,\,f_{19})=1+(f_{19}-1)\; F(E)
\end{align}
with
\begin{equation}  
  \label{eqn:MODIFIER}
  F(E) = \left\{
  \begin{array}{l l}
    \; 0 & \quad E\le1\,\mbox{PeV}\\ \dfrac{\log_{10}(E/1\,{\rm
        PeV})}{\log_{10}(10\,{\rm EeV}/1\,{\rm PeV})} & \quad
    E>1\,{\rm PeV}
  \end{array}\right. \,.
\end{equation}
\begin{figure}[t!]
  \centering 
  \includegraphics[width=.99\linewidth]{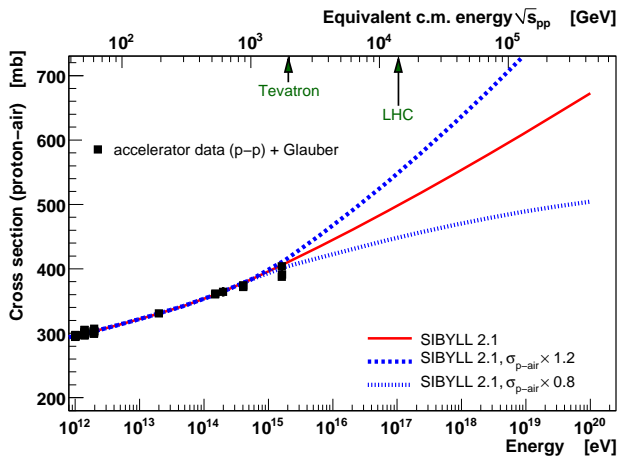}
  \caption{Example of a modified cross section for \textsc{Sibyll} for
    a 20\,\% increase and decrease of $f_{19}$ (see
    \protect\cite{Engel:2000pb,Ulrich:2009yq} for references to the
    data).}
  \label{f:SigmaModifiedCrossSection}
\end{figure}
The factor $f(E,\,f_{19})$ is $1$ below $\unit[10^{15}]{eV}$, where
the models are constraint by accelerator data.  Above
$\unit[10^{15}]{eV}$, the deviation of $f(E,\,f_{19})$ from $1$
increases logarithmically with energy, reaching the value of $f_{19}$
at $\unit[10^{19}]{eV}$, cf.\
Fig.~\ref{f:SigmaModifiedCrossSection}.  This reflects the increasing
uncertainty of the extrapolations with energy.  The factor
$f(E,\,f_{19})$ is then used to re-scale specific characteristic
properties of high energy hadronic interactions, such as the
interaction cross section, secondary particle multiplicity etc..
Obviously, by doing this we rapidly leave the parameter space allowed
by the original model; Thus, for large deviations from the original
model the results are getting less reliable and have to be treated
with some caution. Nevertheless, one can get a clear impression of how
the resulting EAS properties are depending on the specific hadronic interaction
property.

\subsection{Simulation of Showers}

As primary particles we will focus on proton and iron at an energy of
$E_0=10^{19.5}\,$eV. The elements in the mass range from
proton to iron nuclei are the most abundant elements in cosmic
rays. This way our results will bracket the expected mass range of the primary
particles. An energy of about $10^{19.5}\,$eV is the highest energy at
which one can expect to have good statistics of measured air showers
with current and forthcoming observatories \cite{Blumer:2010zz,Santangelo:2009zz}.
Above this energy the flux
of cosmic rays is strongly suppressed
\cite{Abbasi:2007sv,Abraham:2008ru}. 

The simulations are mainly based on the interaction model
\textsc{Sibyll}~\cite{Ahn:2009wx,Engel:1992vf,Fletcher:1994bd}. The choice of this model is not important for our
studies as we will change the model output phenomenologically. It
serves only as a reference for comparison with other models. We also
performed cross checks with other interaction models for proton
primaries.

All shower simulations are performed with the hybrid
code CONEX~\cite{Bergmann:2006yz}.  The CONEX energy threshold above
which particles are tracked individually is set to
$\unit[10^{15}]{eV}$ to require full Monte Carlo simulation for all
interactions above $\unit[10^{15}]{eV}$. The CONEX air shower
simulation program was modified to evaluate $f(E,\,f_{19})$ after each
interaction of a hadronic particle (nucleon or meson) with a nucleus
of air. In case $f(E,\,f_{19})\ne1$ a \emph{resampling}
algorithm~\footnote{Available upon request from
  \texttt{ralf.ulrich@kit.edu}.} is applied to change the properties
of final state particles of the corresponding interaction. 
The resampling algorithm is designed to specifically change only individual
properties of the secondary particle distributions, while conserving
all of the important features like total energy, charge, particle
types, energy fractions in different particle types, and the leading
particle as much as possible.  For the detailed description of
the resampling algorithms see Appendix~\ref{app:mod}.

For primary cosmic ray nuclei with mass number $A>1$, the Glauber
model~\cite{Glauber:1955qq,Glauber:1970jm} is used to relate
nucleus-air interactions to the underlying nucleon-air
interactions. The hadronic event generator \textsc{Sibyll} is based on
the semi-superposition model~\cite{Engel:1992vf}, which allows us to
apply the same resampling algorithms for showers of nuclear
primaries. Thus, for primary cosmic ray nuclei other than protons we
adopt the same modified \textsc{Sibyll} interaction model but with each of
the superimposed nucleon-air interactions being individually changed
according to $f(E,\,f_{19})$--of 
course at the correspondingly lower energy of $E = E_0/A$ compared to
the total energy of the projectile nucleus $E_0$. See
Appendix~\ref{app:modnuc} for details.

\begin{figure*}[pth]
  \centering
  {\includegraphics[width=.47\linewidth]{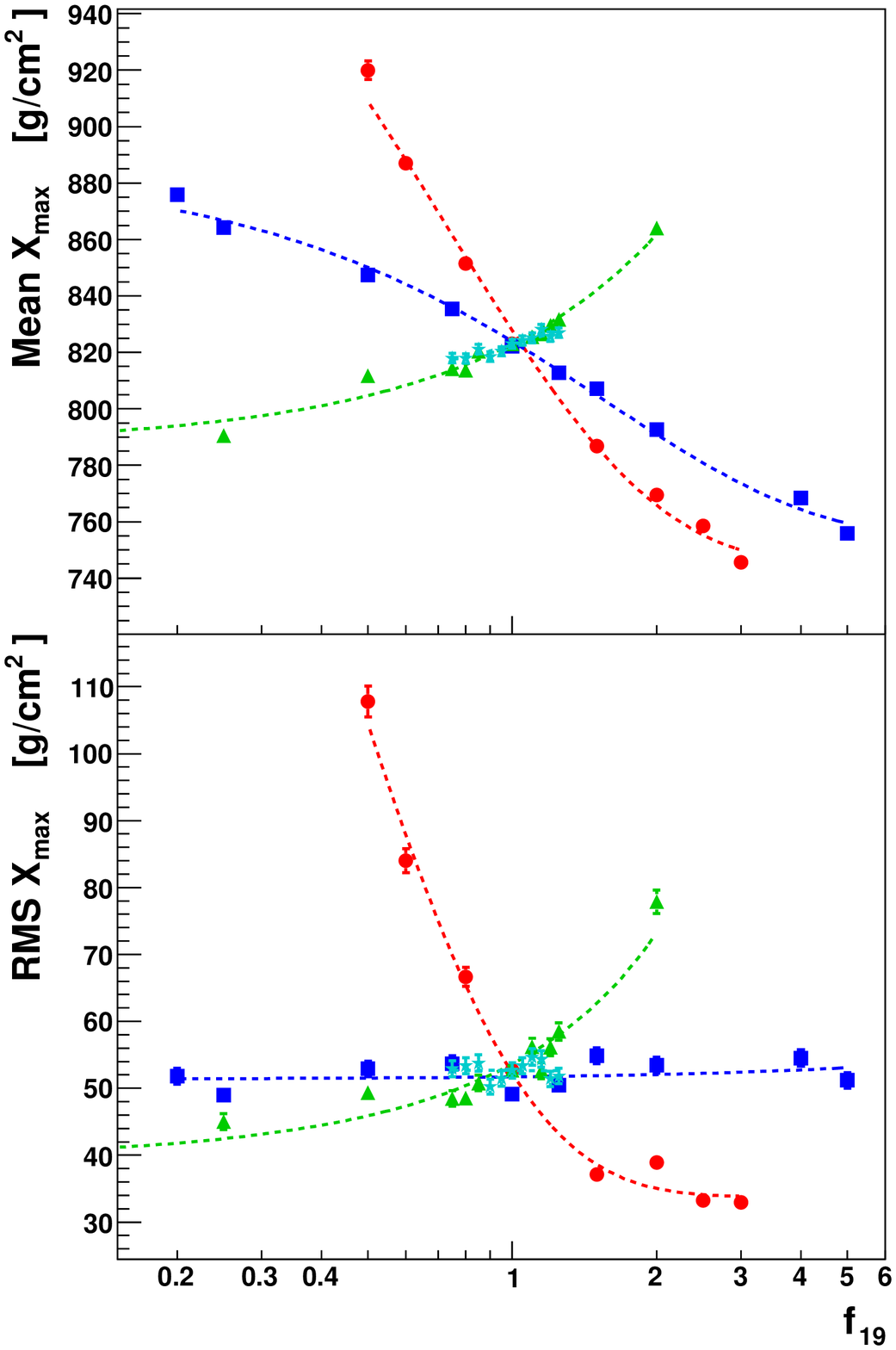}}
  \hspace*{.8cm}
  {\includegraphics[width=.47\linewidth]{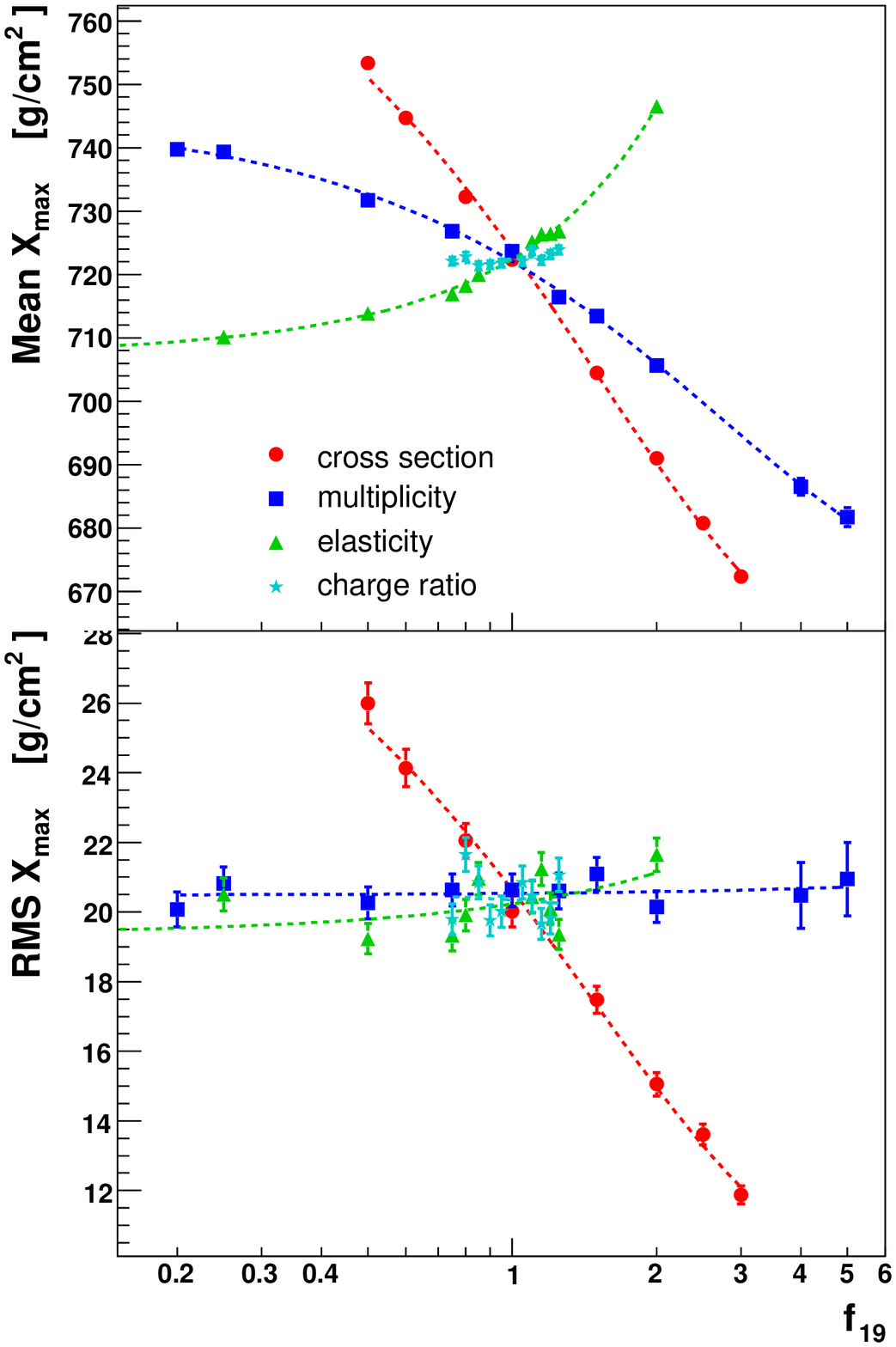}}
  \vspace*{-.4cm}
  \caption{Impact of hadronic interaction features on the shower
    maximum, $X_{\rm max}$, for proton (left) and iron (right) primaries.}
  \label{fig:combXmax}
\end{figure*}

\section{Results}
\label{sec:results}

In the following all simulations are performed for proton and iron
primaries of $\unit[10^{19.5}]{eV}$.  Since the results discussed here are
not very much depending on the particular choice of the primary
energy, the findings are relevant to the analysis of air showers at
least in the energy interval from $10^{19}$ to $10^{20}\,$eV.
For each point in the parameter space under
investigation, 1000 showers are simulated.

In the discussion of our results we will frequently compare to the
analytic Heitler model predictions summarized in Table~\ref{tab:heitler},
and also refer to the dependence of EAS fluctuations on the
longitudinal shower development as shown in Fig.~\ref{fig:protonFluctuations}.

%
\subsection{Longitudinal Shower Development and Depth of the Shower Maximum}
\begin{figure*}[pth]
  \centering
  {\includegraphics[width=.47\linewidth]{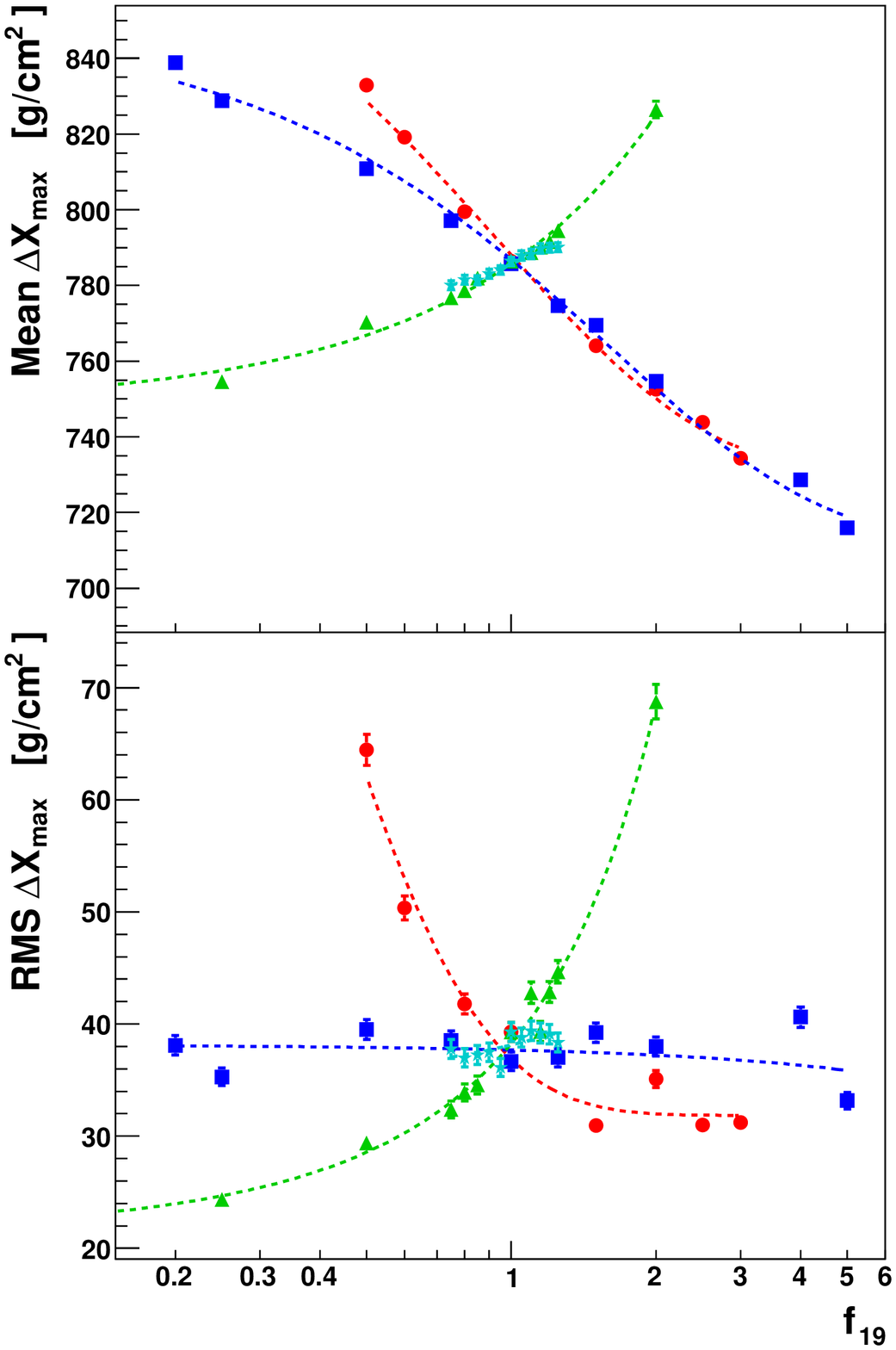}}
  \hspace*{.8cm}
  {\includegraphics[width=.47\linewidth]{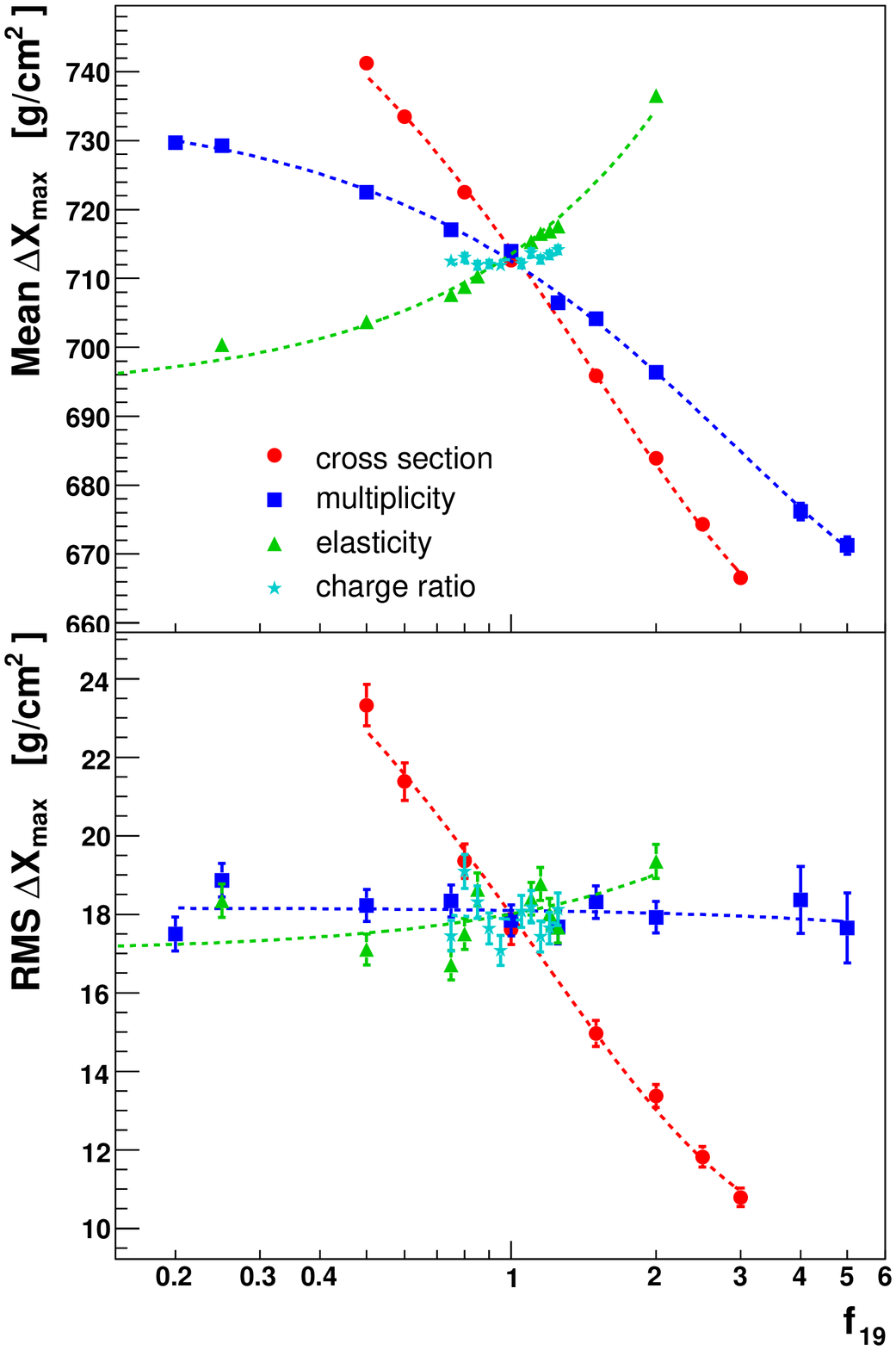}}
  \vspace*{-.4cm}
  \caption{Impact of the modification of hadronic interaction features on $\Delta X$ for proton (left) and iron (right) primaries.}
  \label{fig:combDeltaX}
\end{figure*}
\begin{figure*}[ptbh]
  \centering
  {\includegraphics[width=.47\linewidth]{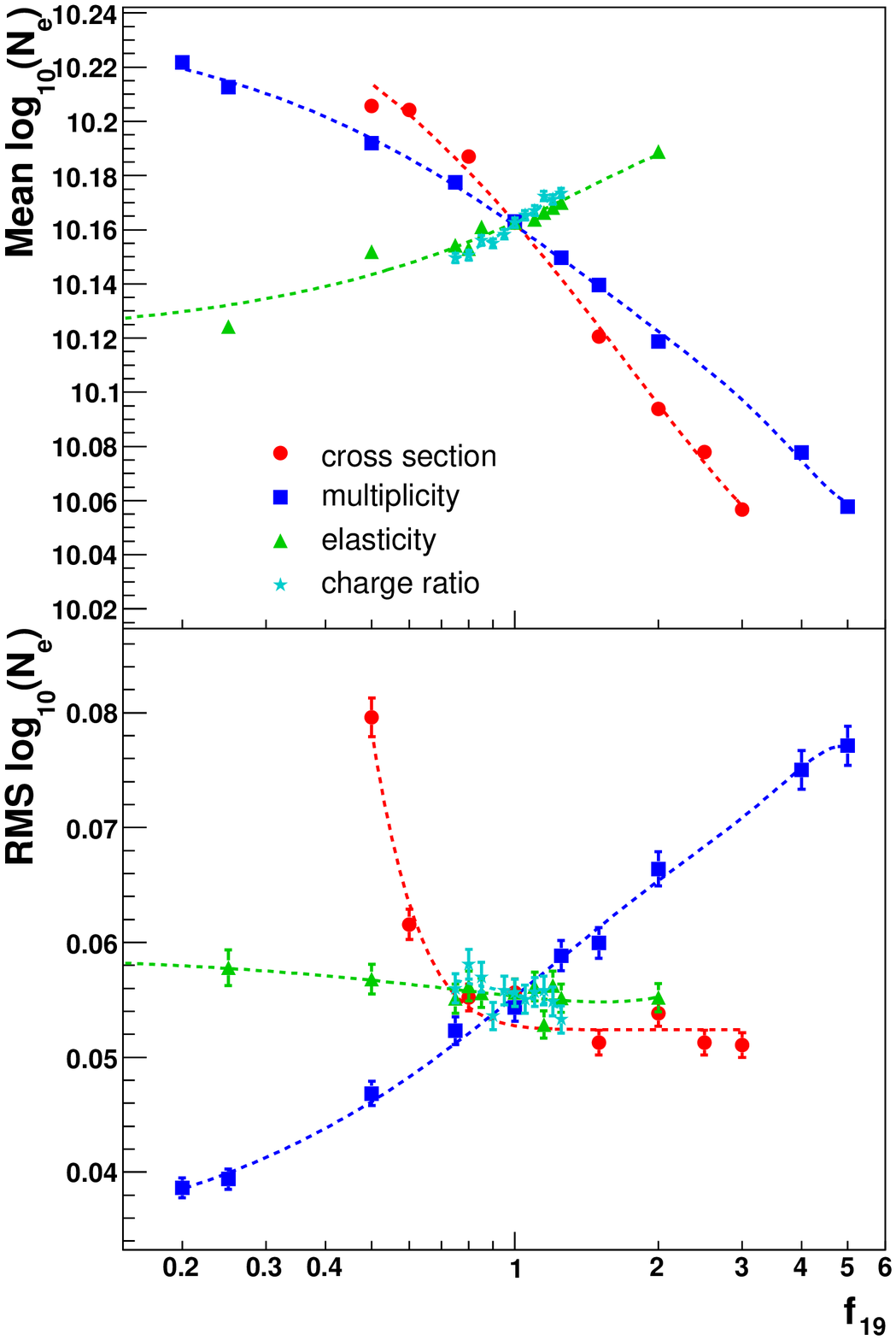}}
  \hspace*{.8cm}
  {\includegraphics[width=.47\linewidth]{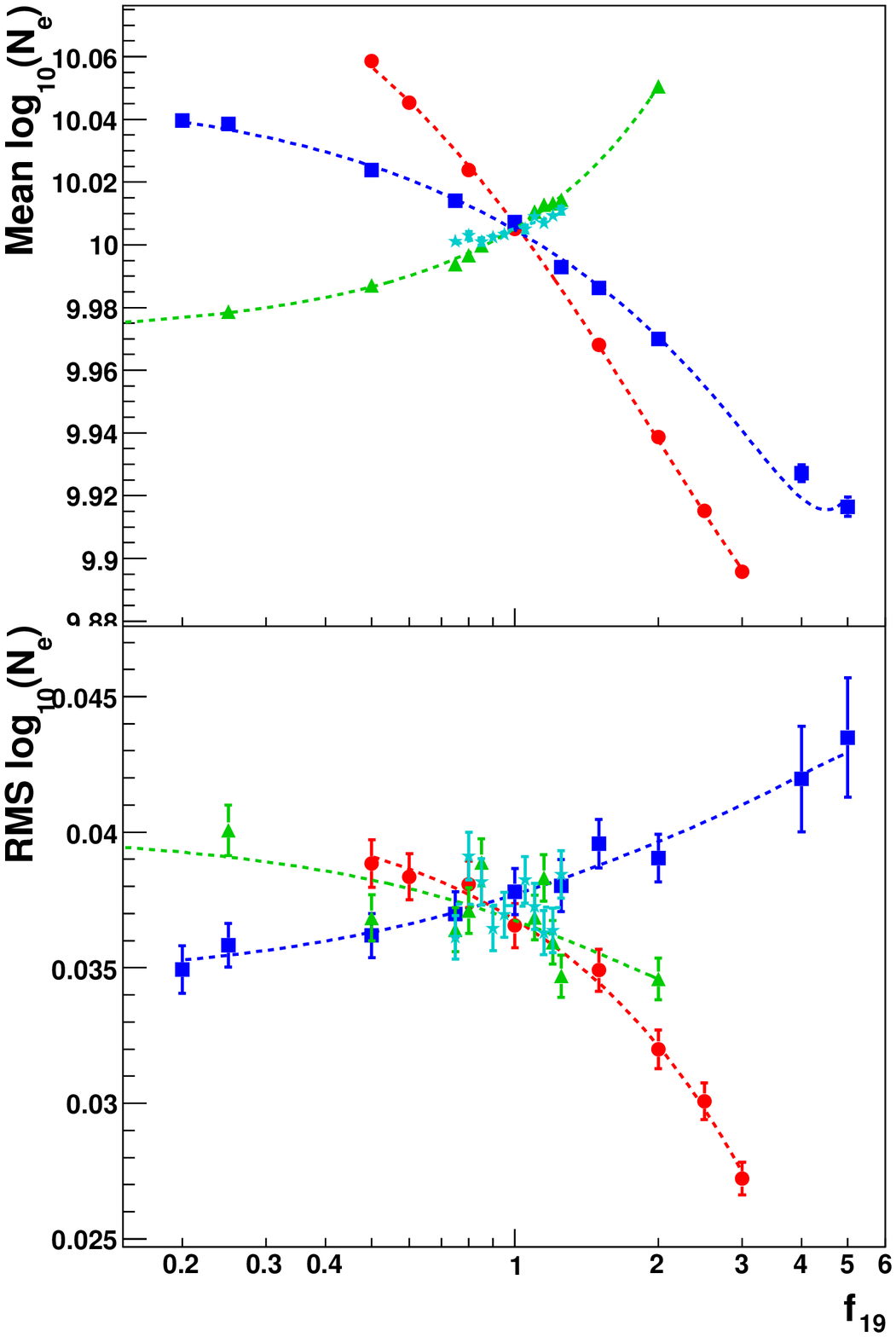}}
  \caption{Impact of a modified extrapolation of hadronic interaction features on the number of
    electrons, $N_{\rm e}$, for proton (left) and iron (right) primaries.}
  \label{fig:combNe}
\end{figure*} 
The results for the mean depth of shower maximum, $\langle X_{\rm
  max}\rangle$, and the fluctuation of $X_{\rm
  max}$, characterized by 
RMS($X_{\rm max}$), are summarized in Fig.~\ref{fig:combXmax}.
The extrapolation of the total cross section for particle production has by far the biggest impact on
$X_{\rm max}$. It can shift $\langle X_{\rm max}\rangle$ by almost
$\unit[100]{g/cm^{2}}$ for protons and $\unit[40]{g/cm^{2}}$ for iron
in both directions, and exhibits a strong correlation with the fluctuations
of $X_{\rm max}$. All the other interaction characteristics considered
here change
the fluctuations only within a few g/cm$^{2}$, except the elasticity for
proton primaries. A high elasticity leads to a
moderate increase in fluctuations, at the same time shifting the
$\langle X_{\rm max}\rangle$ deep into the atmosphere. The secondary
multiplicity is almost as effective in shifting $\langle X_{\rm
  max}\rangle$ as the cross section. This is a consequence of the
distribution of the same energy onto a growing number of particles,
which is also predicted by the Heitler model.  However, the dependence
we find is somewhat different from the simple proportionality to
$-\ln n_{\rm mult}$ for larger deviations from the original model. 
For proton primaries the dependence on the cross section
is similar to $1/\sigma$ 
as in the Heitler model, especially at larger cross sections; 
For iron primaries, on the other hand, this change is more like $-\ln\sigma$.
Furthermore, in contrary to the independence of $\langle X_{\rm
  max}\rangle$ from the pion charge ratio $c$ we find a slight trend
$\propto\ln c$. The impact of the elasticity is
approximately $\propto\kappa_{\rm el}$.

In addition to studying $X_{\rm max}$ we also considered the quantity
$\Delta X=X_{\rm max}-X_1$, with $X_1$ being the depth of the first
interaction in a shower. $\Delta X$ is only sensitive to the shower
development that follows the first interaction.  In
Fig.~\ref{fig:combDeltaX} the results for $\Delta X$ are summarized.

As can be seen, only modifications of the cross section change $X_1$
and $\Delta X$ at the same time. By construction, all other
modifications have an identical impact on $X_{\rm max}$ and $\Delta
X$, since they do not change $X_1$.
Concerning the impact of the cross section extrapolation, we find that
for iron primaries almost the full impact on $X_{\rm max}$ originates
in fact from the air shower development after the first interaction
$\Delta X$. The interaction length of iron in air at
$\unit[10^{19.5}]{eV}$ is just $\unit[\sim9]{g/cm^2}$ while for
protons it is $\unit[\sim38]{g/cm^2}$.  Reducing the cross section by
a factor of $f_{19}=0.5$ at $E=10^{19.5}\,$eV increases the
interaction length by a factor of $1.3$,
which for iron accounts only for about $\unit[\sim12]{g/cm^2}$ of the
total effect of $\unit[\sim35]{g/cm^2}$ seen in $\langle X_{\rm
  max}\rangle$. The rest is caused by the impact of a smaller cross
section on the air shower development, and thus $\Delta X$.
Even for proton primaries the air shower development contributes about
half of the total effect on $\langle X_{\rm max}\rangle$. For example,
lowering the cross section by a factor of 2 increases $\Delta X$ by
$\unit[\sim47]{g/cm^2}$ while $X_{\rm max}$ changes by
$\unit[\sim97]{g/cm^2}$.

For proton primaries the minimal possible fluctuation in $X_{\rm max}$
are about $\unit[\sim35]{g/cm^2}$. This asymptotic behavior of
RMS($X_{\rm max}$) for large cross sections corresponds well to the
predictions of the Heitler model, Eq.~({\ref{eq:heitlerVar}}), where
the asymptotic value of the fluctuations is related to fluctuations of
the secondary multiplicity. For the case of iron primaries no such
saturation effect is observed, which indicates a very much reduced
importance of fluctuations induced by the secondary particle
multiplicity. Also the much smaller overall dependence of RMS($X_{\rm
  max}$) on the cross section for iron induced showers should be noted.

The independence of RMS($X_{\rm max}$) of a modification of the
  multiplicity is related to V[$X_{\rm max}$]$\,\propto\,$V[$\ln n_{\rm
    mult}$] (c.f.\ Eq.~({\ref{eq:heitlerVar}})) and thus (with $f$
  being the rescaling factor)
  \begin{align}
    \label{eqn:xmaxFlutMult}
    {\rm RMS}(X_{\rm
      max})\propto &{\rm V}[\ln(f\cdot n_{\rm mult})]={\rm V}[\ln f]+{\rm V}[\ln
    n_{\rm mult}]\nonumber\\&={\rm V}[\ln n_{\rm mult}]. 
  \end{align}
  A straightforward  rescaling of the multiplicity, as 
    performed by our ad hoc model, does not change the
    $X_{\rm max}$ fluctuations since the relative fluctuations of the
    multiplicity are unchanged.

The results shown in Fig.~\ref{fig:combXmax} can be used to
gain some understanding of possible interpretations of existing
data. For example, 
the fluctuations of $X_{\rm max}$ of $\unit[26\pm
10\;\mathrm{(stat.)}\pm 5\;\mathrm{(syst.)}]{g/cm^2}$ observed by the
Pierre Auger
Collaboration~\cite{Abraham:2010yv}  at
$\unit[10^{19.54}]{eV}$ are compatible at the 1\,$\sigma$ level 
with iron primaries, but also with proton primaries in combination 
with a strong increase of the cross section to higher energies.
Of course, in addition to the
$X_{\rm max}$ fluctuations also the mean depth of shower maximum and
further observables have
to be considered for obtaining a consistent interpretation of the
data. Such an analysis is beyond the scope of this work.

\begin{figure*}[pthb] 
  \centering
  {\includegraphics[width=.47\linewidth]{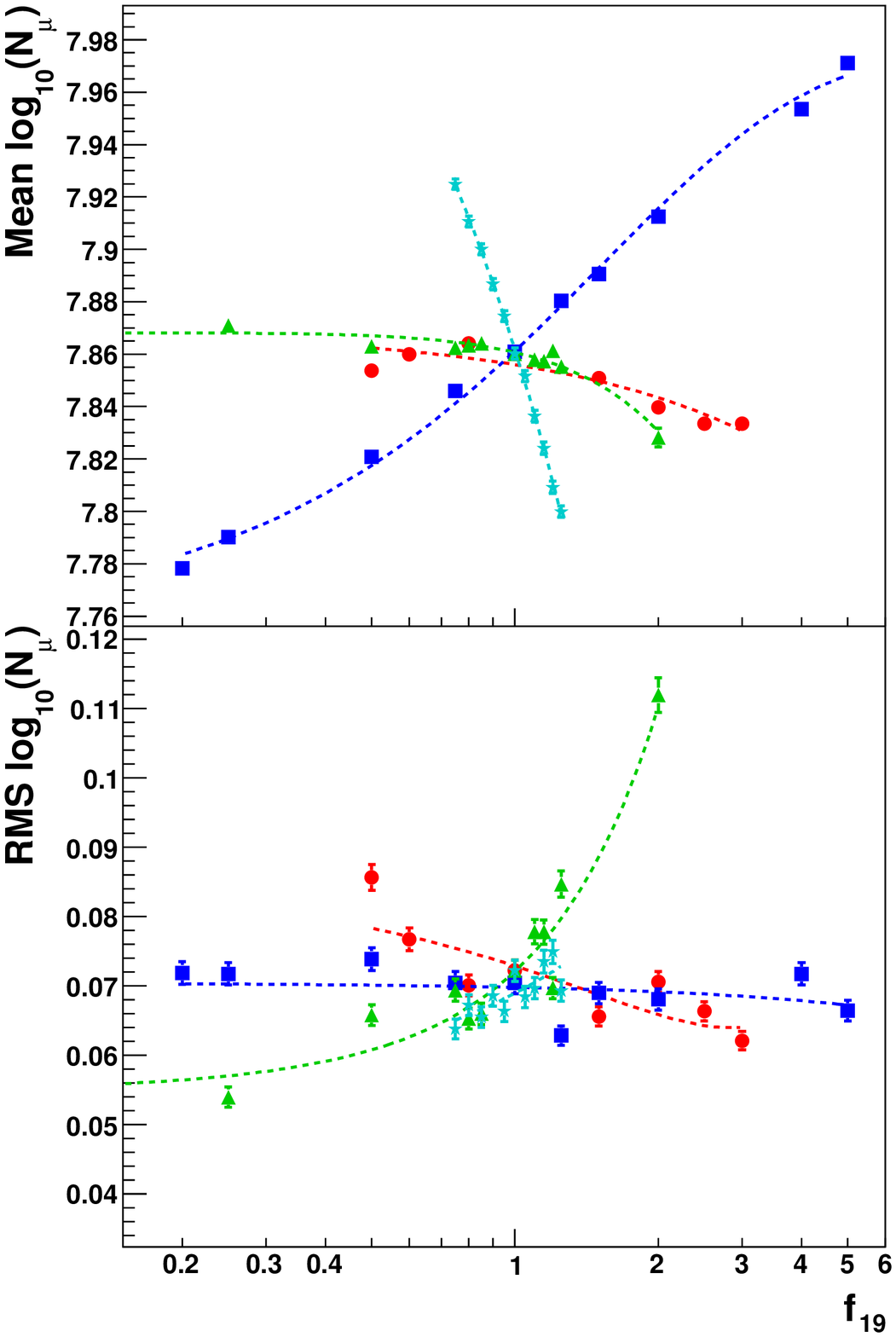}}
  \hspace*{.8cm}
  {\includegraphics[width=.47\linewidth]{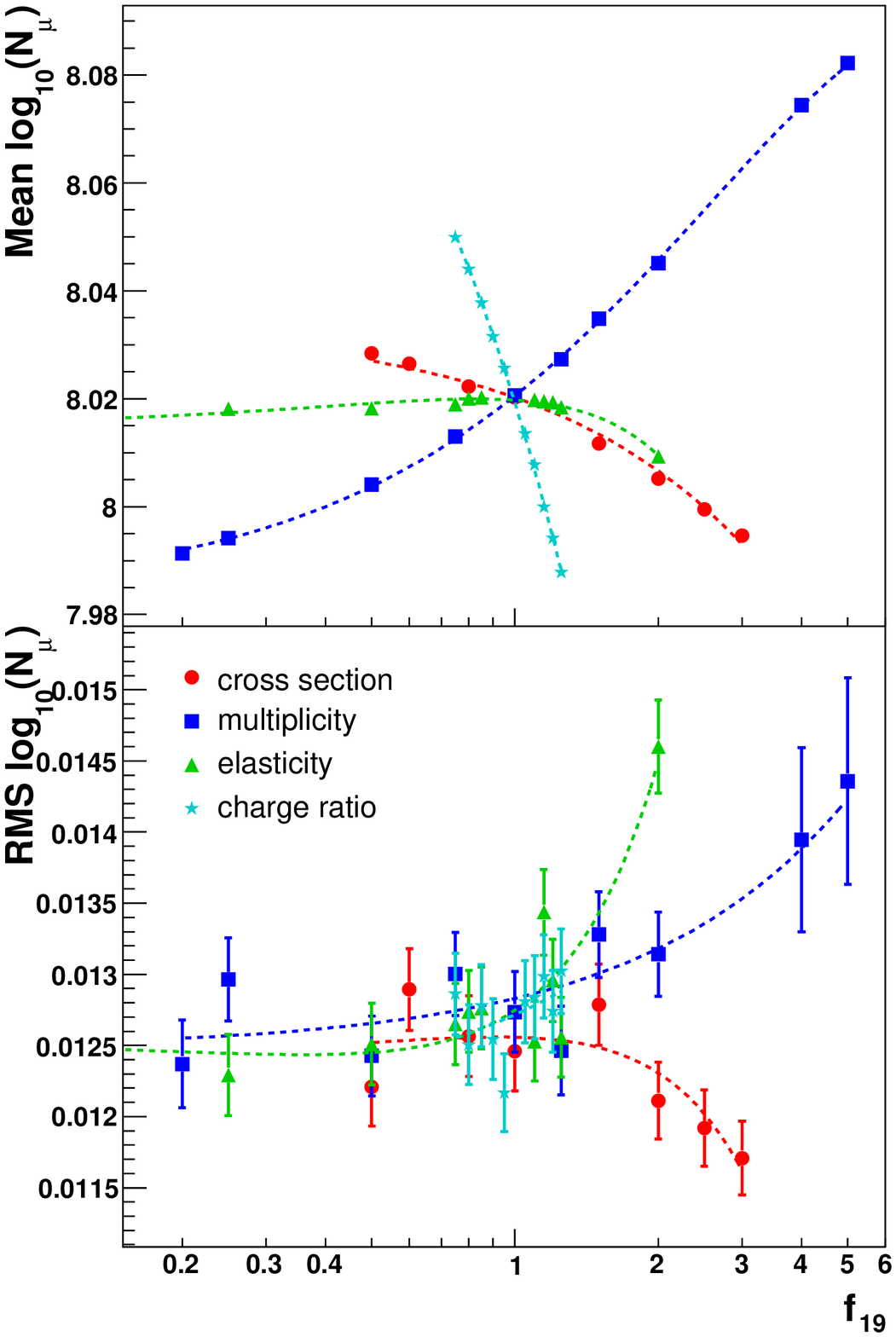}}
  \caption{Impact of a modified extrapolation of hadronic interaction features on the number of
    muons, $N_\mu$, for proton (left) and iron (right) primaries.}
  \label{fig:combNmu}
\end{figure*} 

%
\subsection{Electron Number at \boldmath $X=\unit[1000]{g/cm^{2}}$}
The impact of a modified extrapolation of hadronic interaction
features on the electron number at $\unit[1000]{g/cm^2}$ is shown in
Fig.~\ref{fig:combNe}. The effect on the electron number is twofold.
Firstly, there can be some direct influence of the interaction
characteristics on $\log_{10}N_{\rm e}$.  Secondly, and generally of
even more importance, the change
in the electron number reflects also the attenuation of the shower in
the atmosphere: when $X_{\rm max}$ increases and gets closer to
$\unit[1000]{g/cm^2}$ the number of electrons increases, and when
$X_{\rm max}$ decreases and veers away from the observation level, the
number of electrons decreases. Therefore $\langle\log_{10}N_{\rm
  e}\rangle$ follows very much the corresponding trend in $\langle X_{\rm
  max}\rangle$, cf.~Fig.~\ref{fig:combXmax}.  However, if the impact in
$\log_{10}N_{\rm e}$ would be solely caused indirectly by $X_{\rm
  max}$ then we would e.g.\ expect the fluctuations to strictly follow
the trend shown in Fig.~\ref{fig:protonFluctuations}, which means
fluctuations in $\log_{10}N_{\rm e}$ are minimal at $\langle X_{\rm
  max}\rangle$ and rapidly grow to smaller and larger depths. However,
this trend is qualitatively found only for the modified multiplicity
extrapolation. The elasticity, the charge-ratio and most clearly the
cross section are not showing this behavior, or even an opposite
effect. At the same time we also find that the influence of the
multiplicity on $\langle\log_{10}N_{\rm e}\rangle$ is largest, and the
importance of the extrapolation of the cross section, the elasticity,
and the charge-ratio are reduced compared to that found for $X_{\rm
  max}$. All this is evidence that in addition to the indirect
correlation by the longitudinal EAS development, the changes in the
overall particle production within the air shower have a strong direct
impact on the electron number. See Eq.~(\ref{eqn:NeHeitler}) to get a
qualitative description of how $N_{\rm e,\,max}$ and $X_{\rm max}$ are
both affecting the electron number at ground level, $N_{\rm e}$. Note,
that in Fig.~\ref{fig:combNe} the limit $X_{\rm obs}\gg X_{\rm max}$
is not always fulfilled.

It is interesting to note, that an efficient way to change the
fluctuations of the electron number in proton showers is via the
extrapolation of the multiplicity, especially in the directions of
reduced fluctuations. To increase the fluctuations, lowering the cross
section can be equally effective. For iron primaries increasing the
cross section leads to decreased fluctuations, while the overall
impact of the multiplicity is very much reduced.

\begin{figure*}[thbp] 
  \centering
  {\includegraphics[width=.47\linewidth]{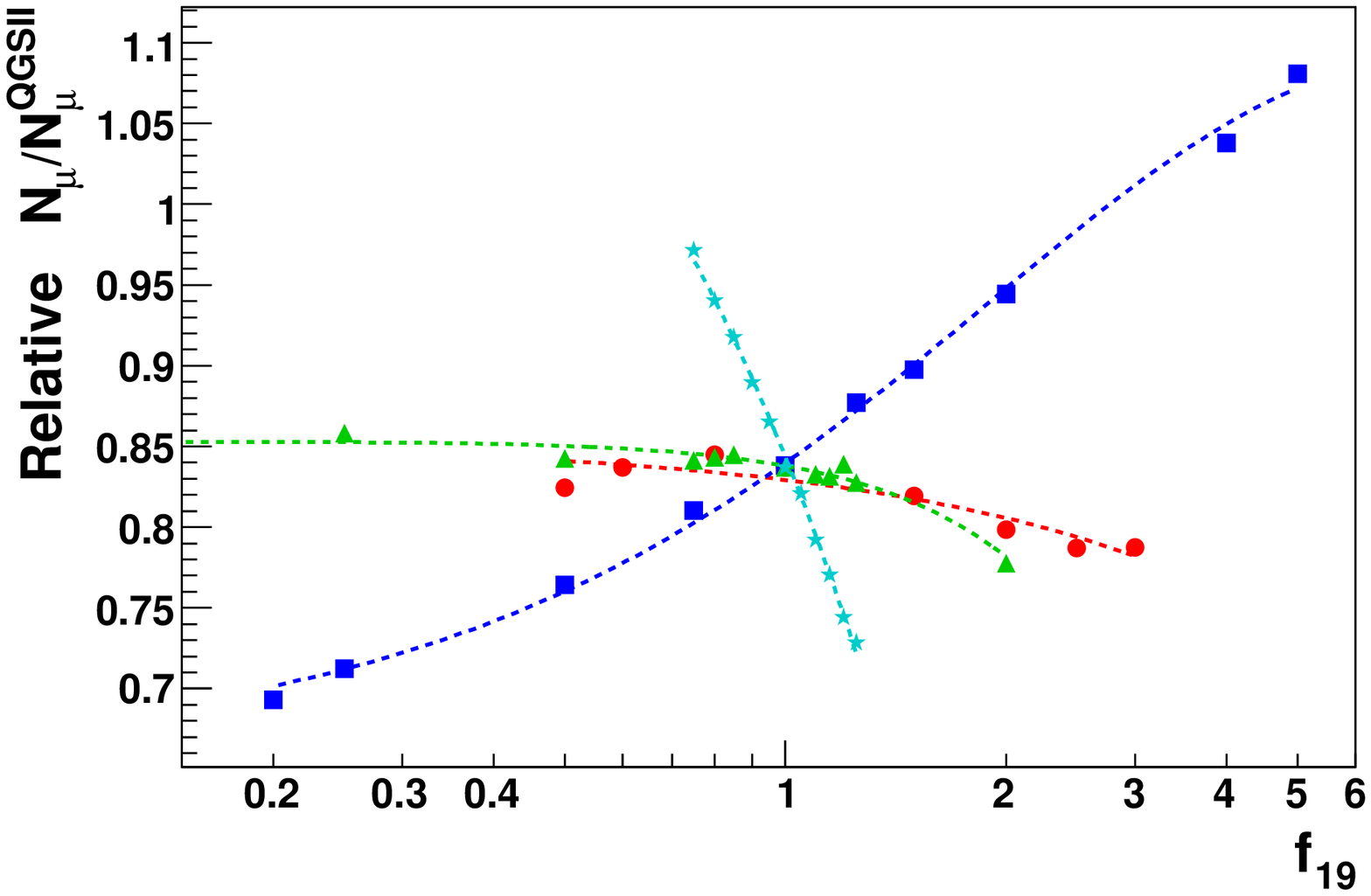}}
  \hspace*{.8cm}
  {\includegraphics[width=.47\linewidth]{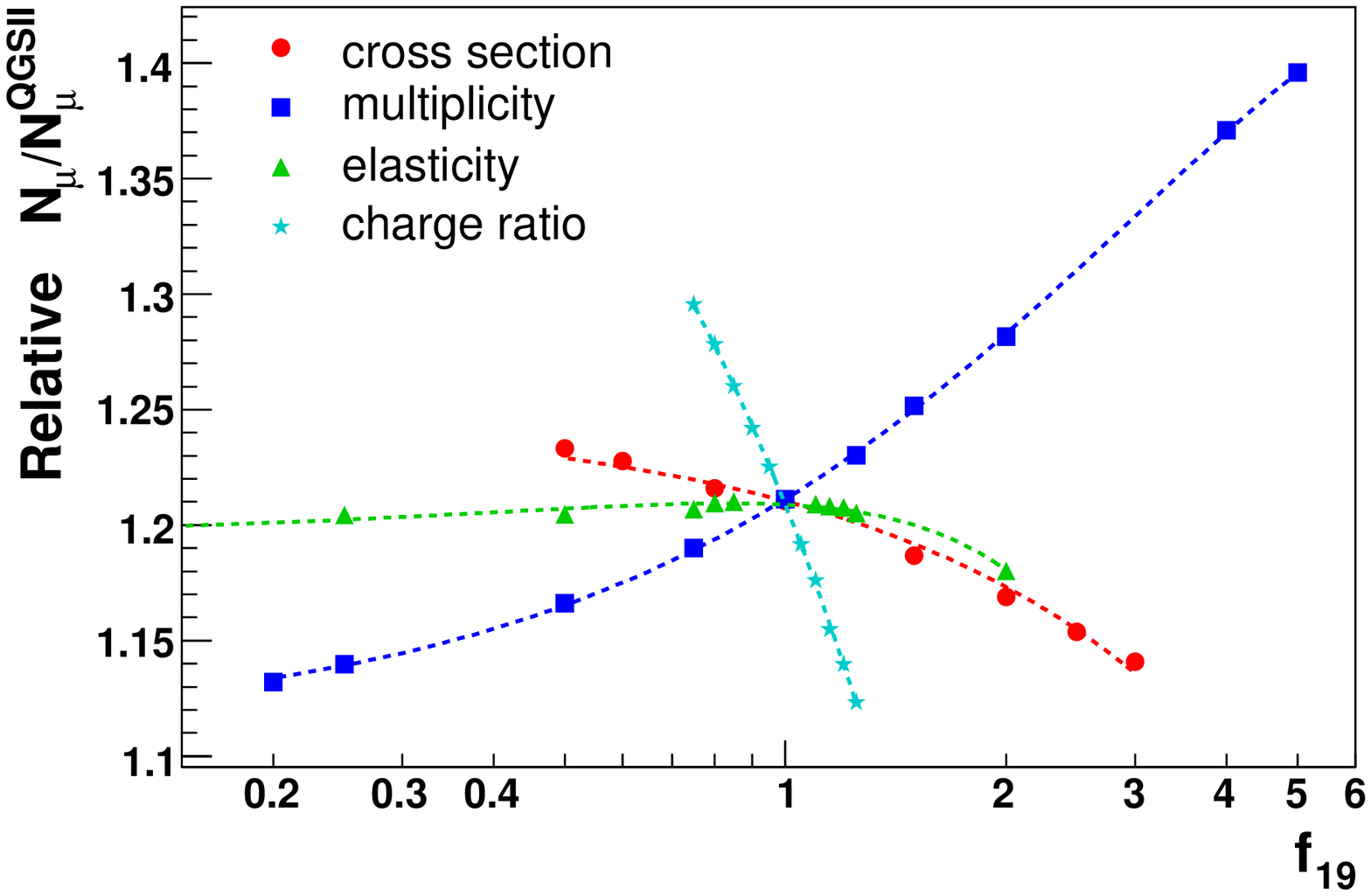}}
  \caption{Influence of the modification of hadronic interaction features on the number of muons relative to the prediction of protons/\textsc{QGSJetII} for proton (left) and iron (right) cosmic ray primaries.}
  \label{fig:combNmuRel}
\end{figure*} 

%
\subsection{Muon Number at \boldmath $X=\unit[1000]{g/cm^{2}}$}
The results of the influence of the modification of interaction features on the muon number
are summarized in Fig.~\ref{fig:combNmu}.
In analogy to electrons also the muon number at $\unit[1000]{g/cm^2}$
reacts to changes in the depth of $X_{\rm max}$ relative to the
observation level. But as shown in Fig.~\ref{fig:protonFluctuations}, this
sensitivity to $X_{\rm max}$ is much smaller than in the case of electrons. 
Especially the fluctuations are not having the clear minimum at
$\langle X_{\rm max}\rangle$, but show a rather smooth transition to
a very constant rate of fluctuations at larger depths. 
Furthermore, since muons in air showers are mainly produced via the
decay of pions, their abundance is very sensitive to the overall
number of pions in the shower. During the shower development there is
a competition between pion decay, yielding muons and neutrinos, and
interaction, producing new hadronic secondaries.
While the hadronic interaction length of pions is $\lambda_{\rm
  int}\sim\unit[120]{g/cm^2}$ over a wide range in energy, the decay
length changes with energy ($\gamma=E/m$) and is $\lambda_{\rm dec}=\rho {\rm
  c}\gamma \tau_{\rm dec}$.
 Muons are produced mainly by pions with $\lambda_{\rm dec}<\lambda_{\rm int}$. This is, for
example, why high energy muons are more efficiently produced at
high altitude where $\rho$ is small. When muons are
produced, pions are removed from the multiplicative hadronic shower
cascade and ultimately fewer pions in the air shower lead to the
production of fewer muons. In addition,  an increase of the production of
high-energy muons at high altitude 
reduces the production of the more abundant muons at lower energies.

Since muons are hardly attenuated in the atmosphere
(cf. Fig.~\ref{fig:protonFluctuations}) the correlation of the muon
number with the location of $X_{\rm max}$ is much reduced in
comparison to the electron number. For muons the actual particle
production characteristics in the EAS are much more directly
important. This is demonstrated by the large \emph{positive}
correlation of $\langle\log_{10}N_\mu\rangle$ with the multiplicity, while
for electrons a \emph{negative} correlation is observed. In our
simulations the muon number increases almost proportional to $\ln
n_{\rm mult}$. The Heitler model predicts a smaller effect of just
$-1/\ln n_{\rm mult}$ at large multiplicities.

The muon number is also very sensitive to the charge-ratio. A change
in the charge-ratio is directly leading to a shift of the fractions of
the primary 
energy going into the muonic and the electromagnetic shower
components. The Heitler model indicates $\log_{10}N_\mu\propto -c$ and our
results are consistent with this.

Similar to the case for the electron numbers the prediction by the
Heitler model for the dependence on the elasticity is
$\log_{10}N_\mu\propto\kappa_{\rm el}$, but our simulations show an opposite effect more
like $\propto-\kappa_{\rm el}$. The fluctuation in $\log_{10}N_\mu$ for proton
primaries depend mainly on the
elasticity.

In Fig.~\ref{fig:combNmuRel} we show the change of the muon number
obtained with the modified version of \textsc{Sibyll} relative to the
prediction of \textsc{QGSJetII}. Within the
\textsc{Sibyll} model, the central parameters of the interaction have
to be changed substantially to predict
the same or even more muons than in \textsc{QGSJetII}. Only extreme
modifications of the multiplicity or the charge ratio lead to an
increase the muon number to the level of \textsc{QGSJetII}. However,
one has to keep in mind that \textsc{Sibyll} is the model with the
smallest muon numbers.

For example \textsc{Epos~1.61}, which
predicts the highest muon numbers, yields up to $\sim70$\,\% more
muons than \textsc{Sibyll}. This difference cannot be explained by a
modification of the interaction features discussed here, whose change would
lead to an increase of the muon number by only
$\sim30\,\%$.  As noted by the authors of \textsc{Epos}, a larger number of
baryon-antibaryon pairs produced in this model keeps a larger fraction
of the primary energy in the multiplicative hadronic shower component to eventually
produce a significantly larger number of pions and thus muons at low
energy~\cite{Pierog:2006qv}. Another important aspect is the chance probability of
producing a neutral pion as a leading particle in charged pion
interactions with air nuclei~\cite{Drescher:2007hc}. Differences in
these features of hadronic interactions lead to large differences in
the total number of muons predicted for air showers.



\begin{figure*}[tbh!]
  \centering
  {\includegraphics[width=.47\linewidth]{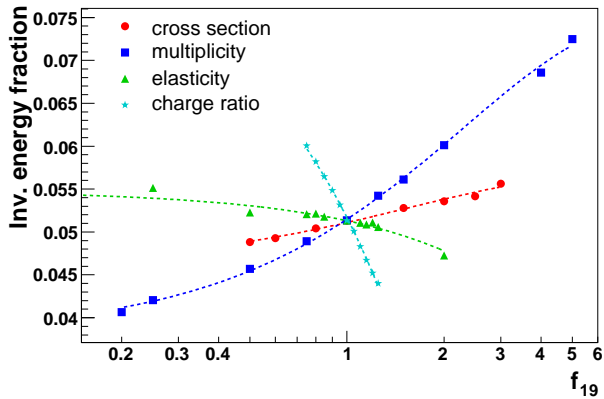}}
  \hspace*{.8cm}
  {\includegraphics[width=.47\linewidth]{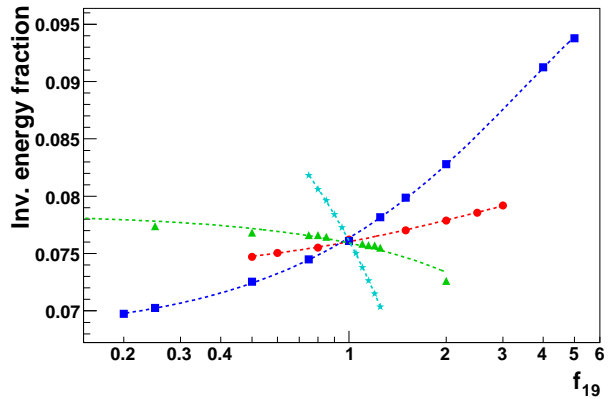}}
  \caption{Impact of modified extrapolations of interaction features on the invisible
    energy fraction, $\epsilon_{\rm inv}$, for proton (left) and iron (right) primaries.}
  \label{fig:combInvE}
\end{figure*}

Recently the Pierre Auger Collaboration reported a measurement of the
density of muons in showers of $\unit[10^{19}]{eV}$ at
$\unit[1000]{m}$ from the core. Using proton induced showers as a
reference, the number of muons is found to be about $1.3$ times higher
than that predicted by using \textsc{QGSJetII} for high energy and
\textsc{FLUKA} for low energy
interactions~\cite{Abraham:2009ds}. Our studies show that increasing the total muon number
in showers by 30\,\% requires, for a given primary composition, rather
drastic changes of the features of the hadronic interactions
considered in this work. In contrast, already moderate modifications
of the modelling of low energy interactions lead to a change of the
muon number of this order, see discussion
in~\cite{Drescher:2003gh,Meurer:2005dt,Maris:2009uc}. A good knowledge
of particle production is needed both at high and low energy to be
able to use the muon number of showers for absolute measurements of
the mass of the primary particle.

%
\subsection{Invisible Energy}
It is one of the advantages of the observation of air showers by
fluorescence telescopes that the measured signal is proportional to
the energy deposit ${\rm d}E/{\rm d}X$ of the air shower in the
atmosphere~\cite{Song:1999wq,Barbosa:2003dc,Nerling:2006yt}. The integral
\begin{equation}
  E^{\rm e.m.}_0 = \int {\rm d}X\; \frac{{\rm d}E}{{\rm d}X}(X) = \int {\rm d}X\; \alpha_{\rm eff}(X) N_{\rm e}(X)\;,
\end{equation}
where $\alpha_{\rm eff}$ is the effective (i.e.\ average) energy loss per
electron, yields then an accurate
measurement of the total energy of the primary cosmic ray particle
that was transferred to the electromagnetic shower component.
Muons and hadrons are much less abundant compared to electrons and also ionize less
efficiently, and neutrinos do not deposit any energy. The
\emph{invisible energy fraction}~\cite{Barbosa:2003dc} is
\begin{eqnarray}
  \epsilon_{\rm inv} &=& (E_0 - E^{\rm e.m.}_0) / E_0 \qquad\quad \text{and thus}\nonumber\\
  E_0 &=& E_0^{\rm e.m.} / (1 - \epsilon_{\rm inv}) 
\end{eqnarray}
cannot be calculated in a purely model independent way, since
$\epsilon_{\rm inv}$ has to be estimated from air shower
simulations. Fortunately $\epsilon_{\rm inv}$ is only of the order of
a few percent (see Fig.~\ref{fig:combInvE}) and thus the introduced
model dependence of the estimated total shower energy is small.

The invisible energy fraction is
naturally correlated to the number of muons in the air shower. The
trends we observe are indeed very similar to what is seen for muon
numbers.
However, the impact on $\epsilon_{\rm inv}$ is generally larger than
for $N_\mu$, indicating the importance of additional effects like
enhanced muon production and differing energy distribution of
electrons or muons. Most strikingly the small trend to smaller muon
numbers of a rising cross section is reversed into a trend to larger
$\epsilon_{\rm inv}$.  
A large cross section causes an accelerated shower development.  The rise of the invisible
energy is related to an increasing number of high energy muons
produced high up in the atmosphere.

Our simulations indicate that, over the full parameter range of the
modifications, the invisible energy fraction does not
change by more than $\sim0.02$.

\section{Dependence on the Interaction Model}
\label{sec:modelDependence}
\begin{figure*}[p]
  \centering
  \includegraphics[width=.47\linewidth]{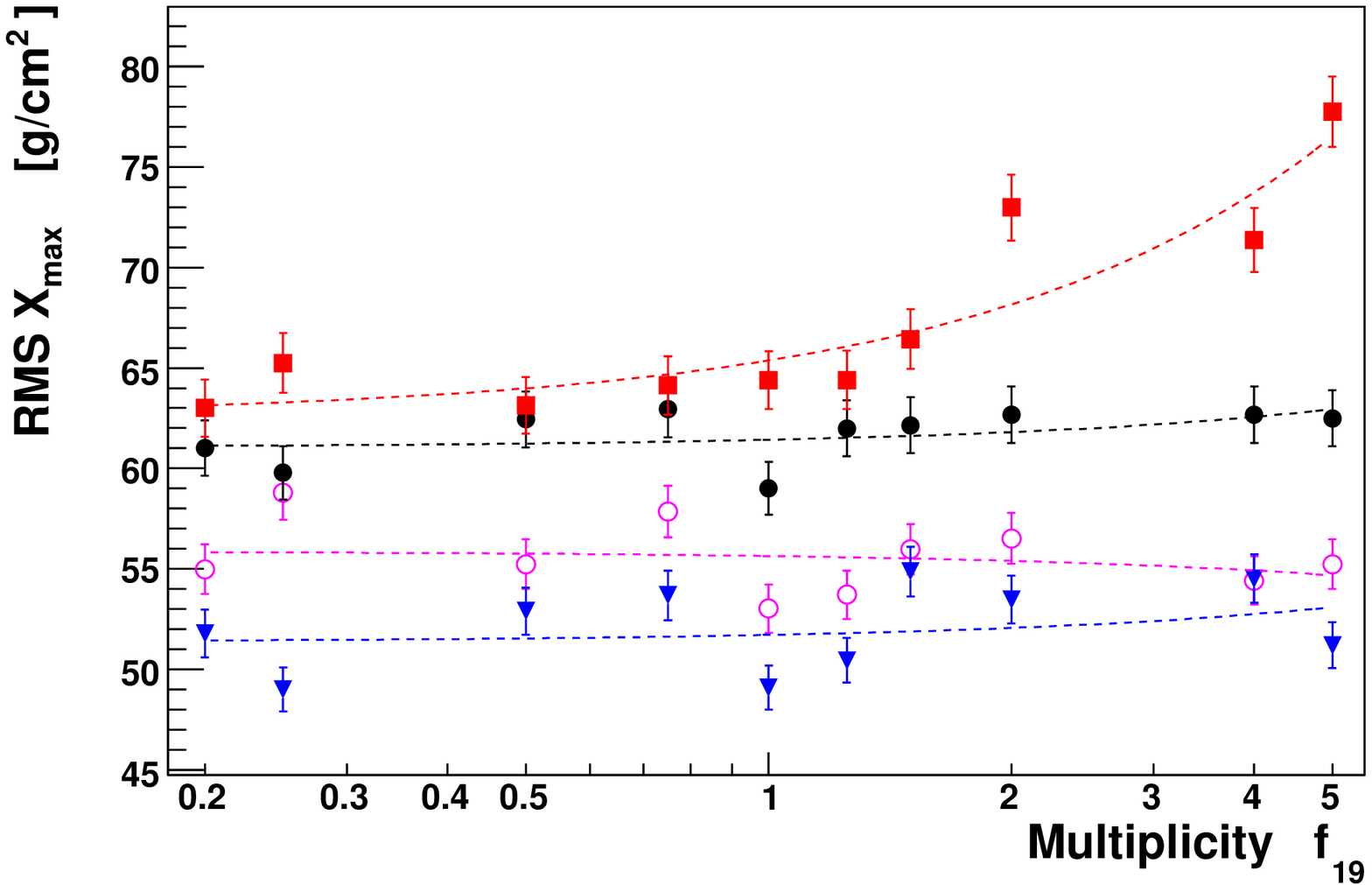}~
  \includegraphics[width=.47\linewidth]{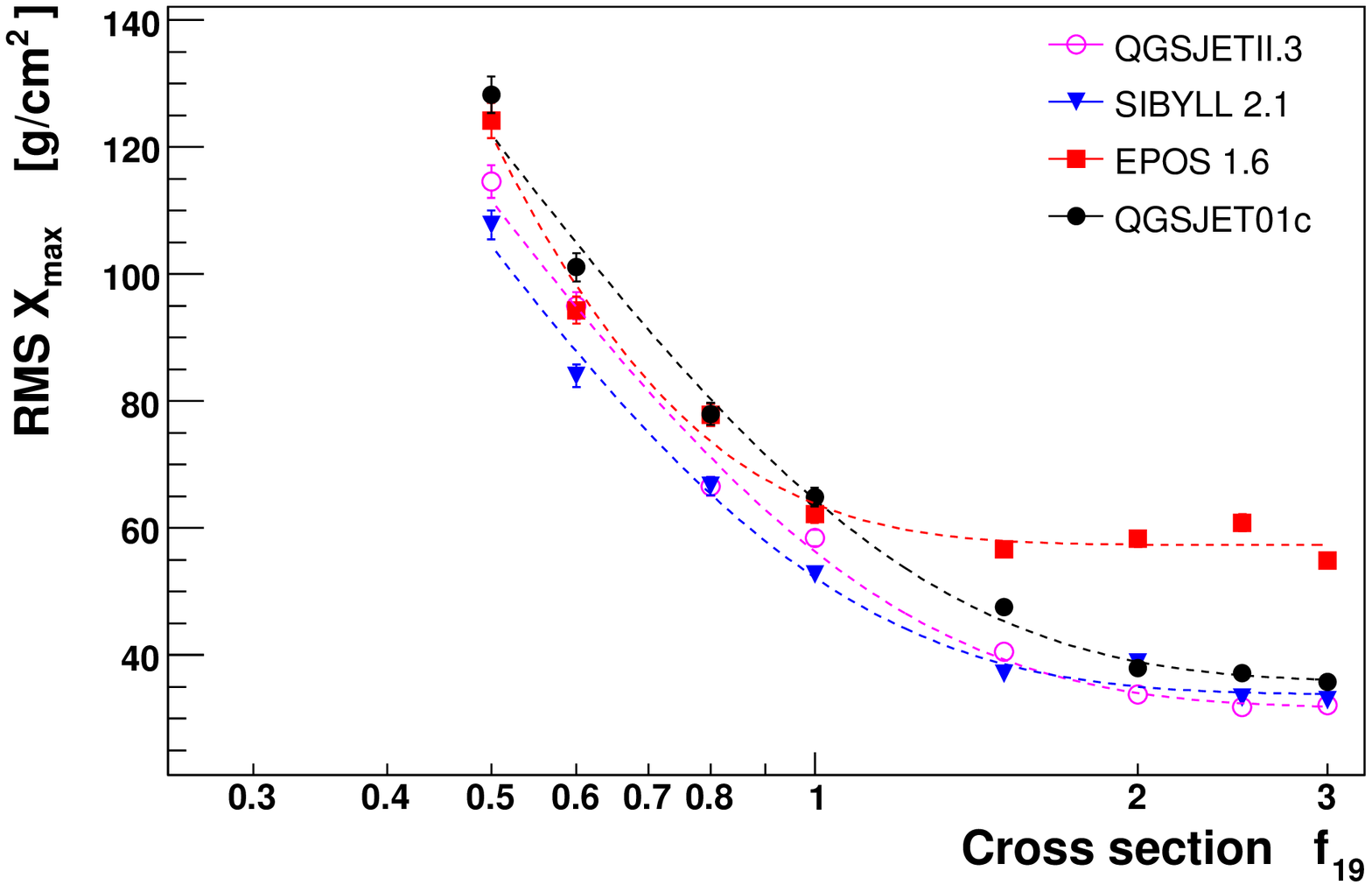}\\
  \includegraphics[width=.47\linewidth]{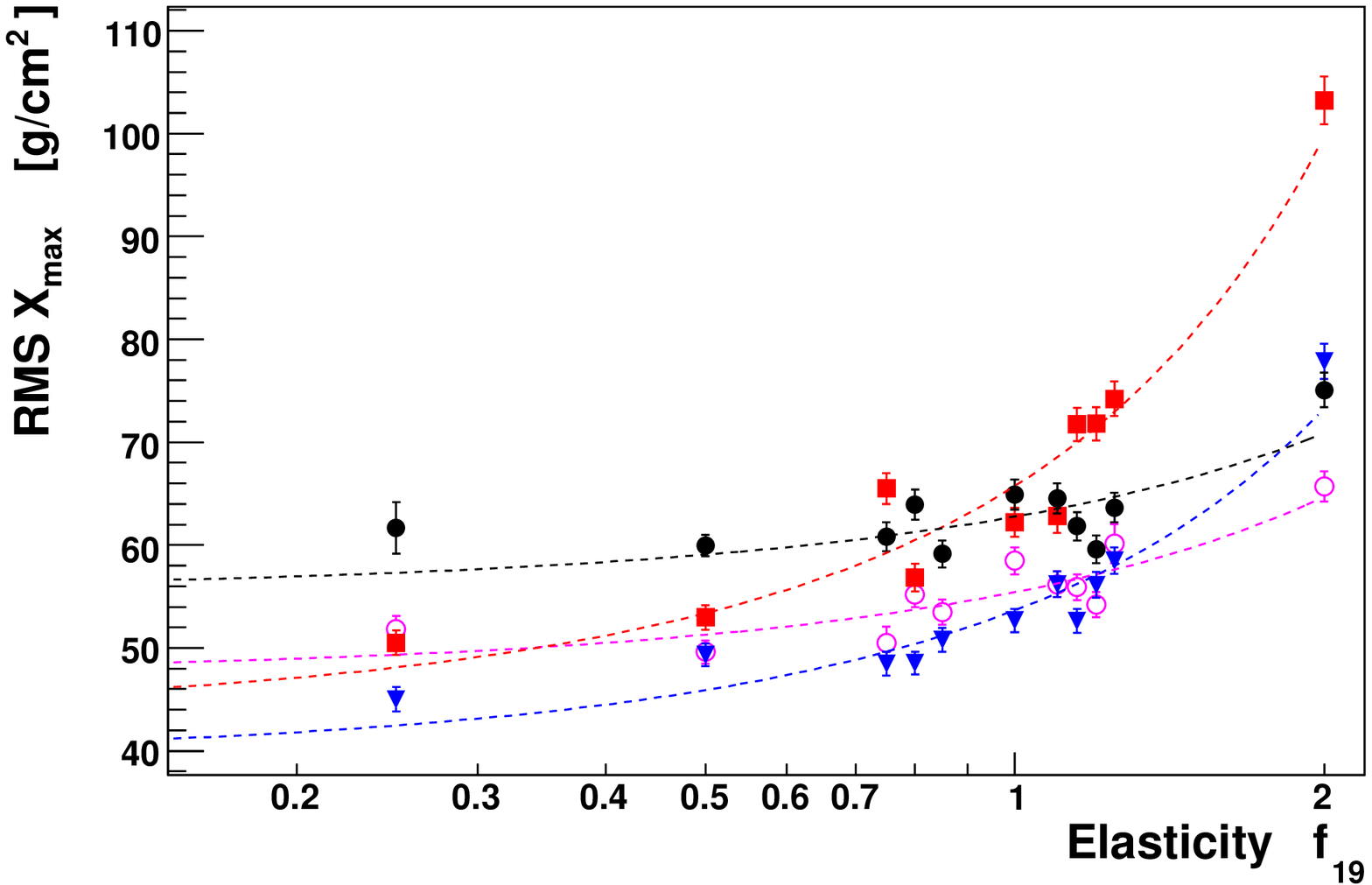}~
  \includegraphics[width=.47\linewidth]{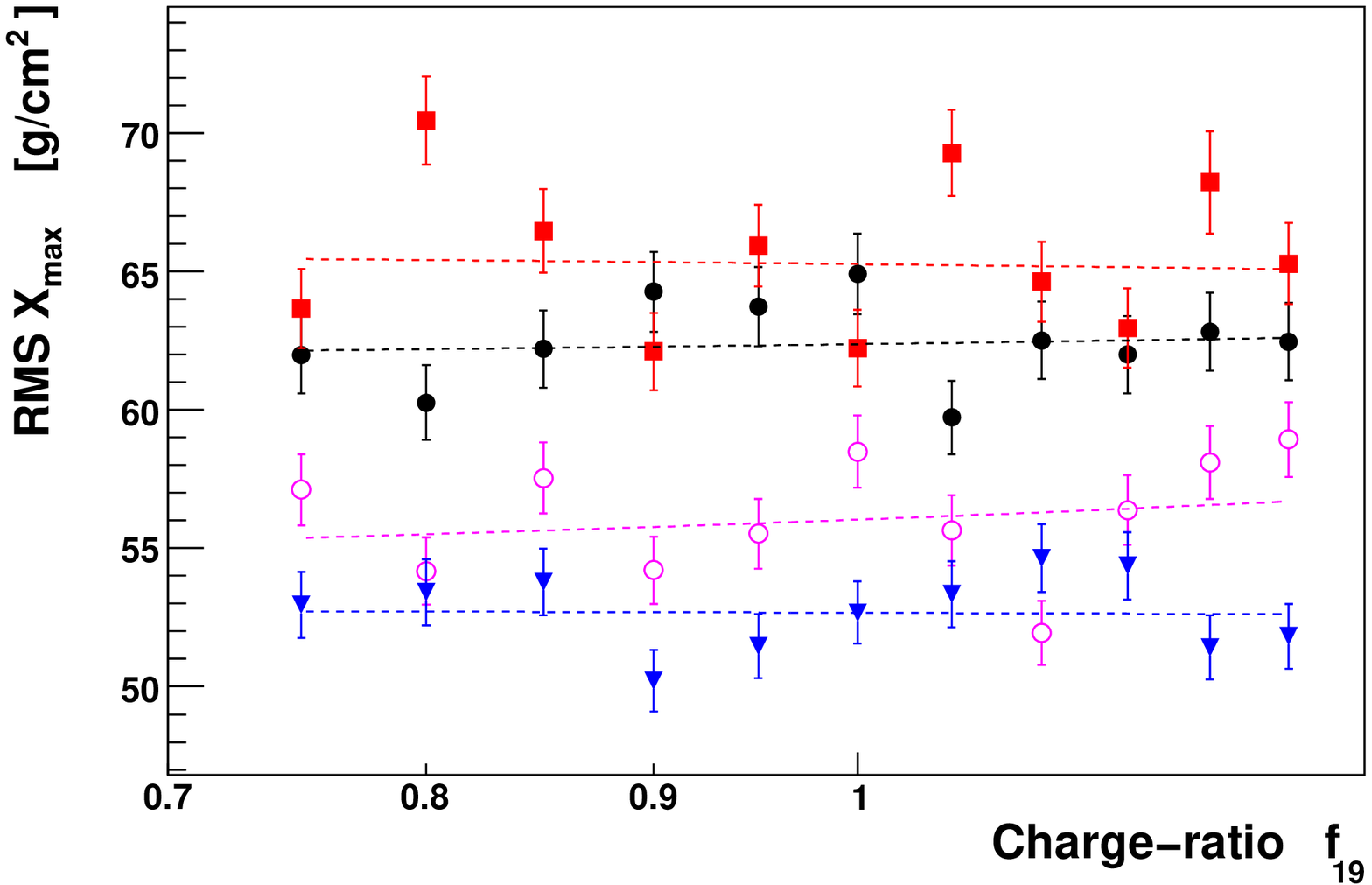}
  \vspace*{-.2cm}
  \caption{Influence of modified extrapolations of particle production
    on RMS($X_{\rm max}$) for several
    hadronic interaction models.}
  \label{modelDepXmaxRMS}
\end{figure*}
\begin{figure*}[p]
  \centering
  \includegraphics[width=.47\linewidth]{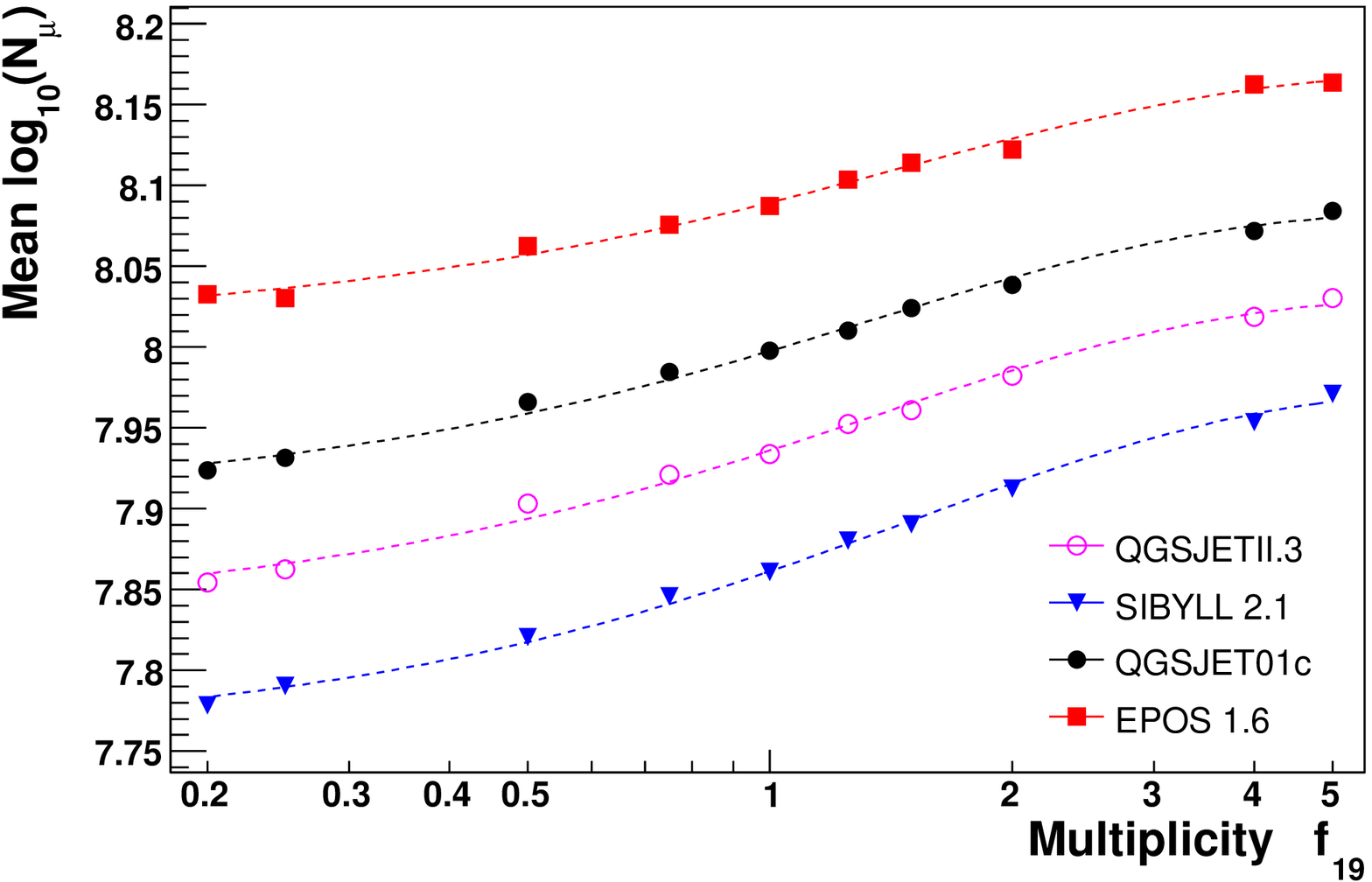}~
  \includegraphics[width=.47\linewidth]{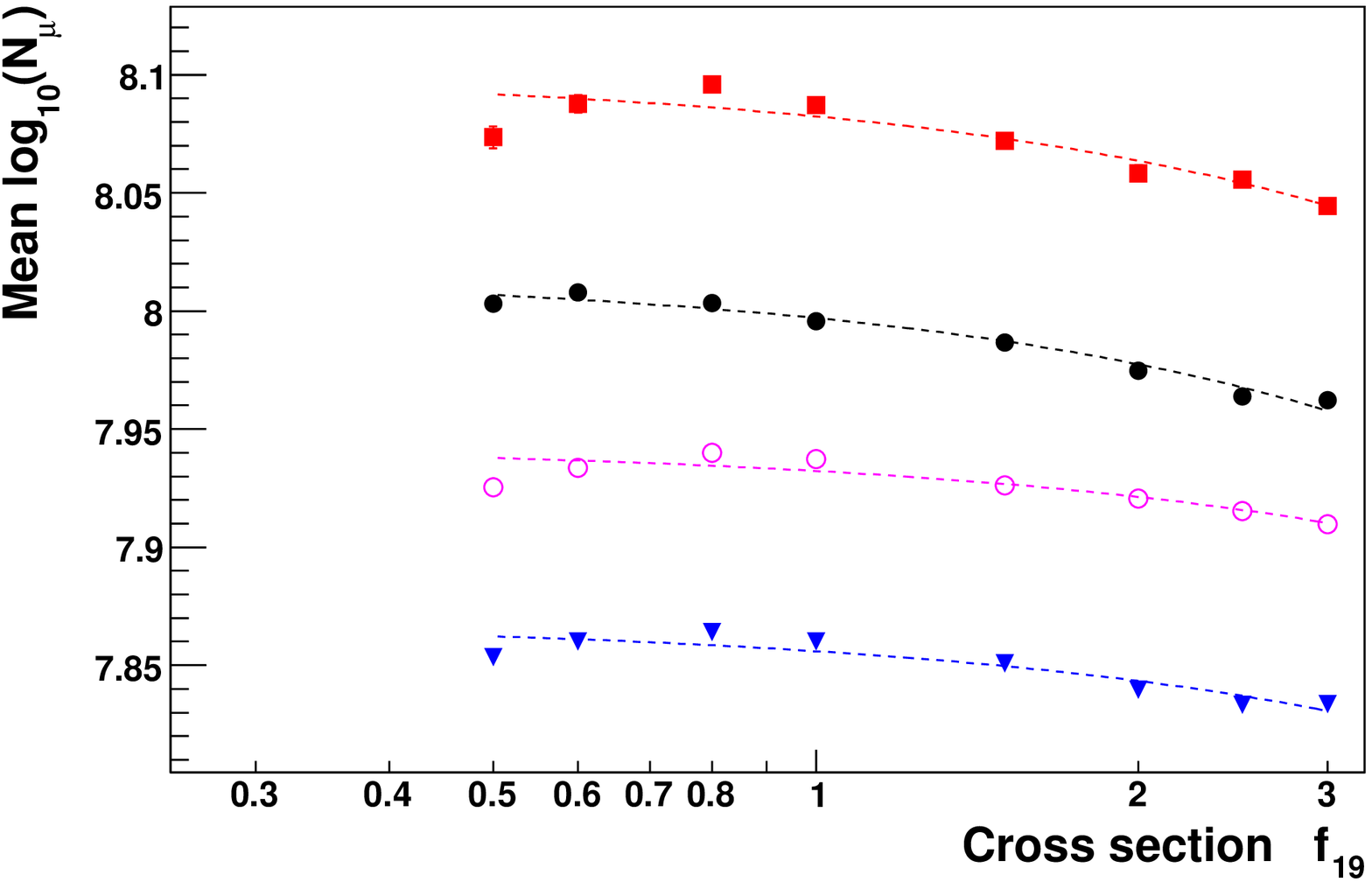}\\
  \includegraphics[width=.47\linewidth]{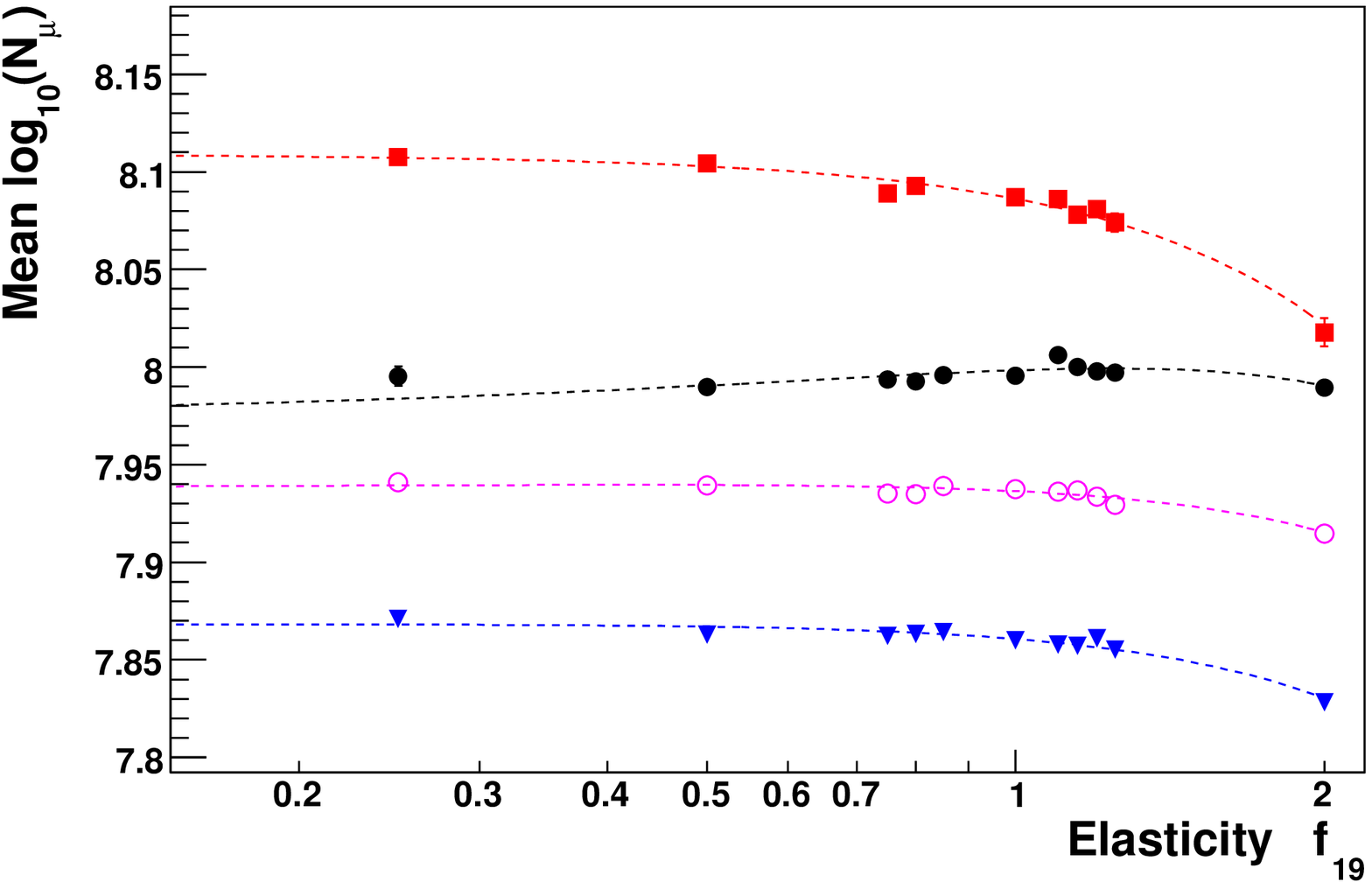}~
  \includegraphics[width=.47\linewidth]{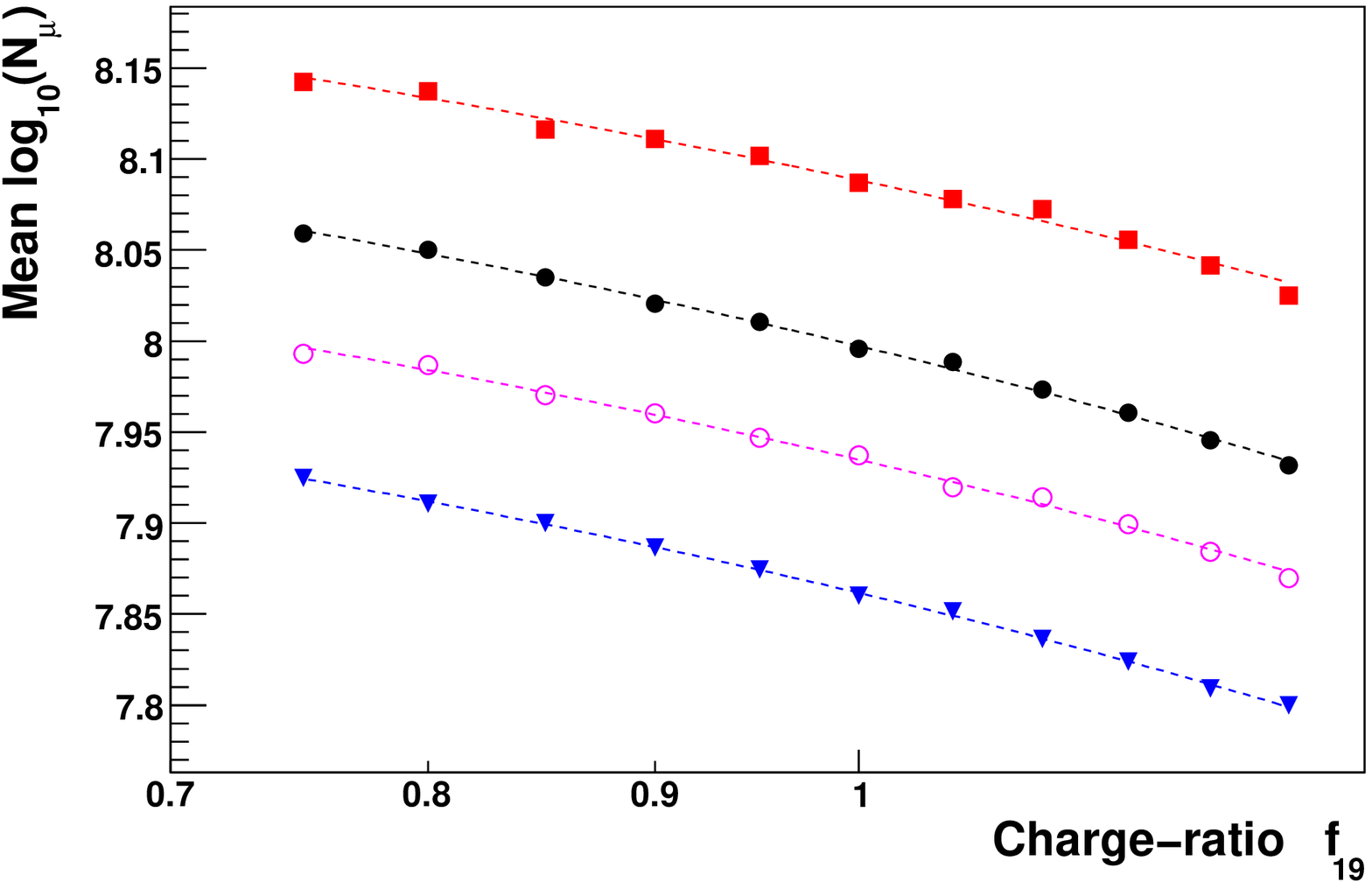}
  \vspace*{-.2cm}
  \caption{Impact of modifications of the extrapolation of particle
    production on $\langle \log_{10}N_\mu\rangle$
    for several hadronic interaction models.}
  \label{modelDepNmuMean}
\end{figure*}
So far all the calculations  were based on \textsc{Sibyll~2.1}. We will now
present a comparison of the previously discussed results with that
obtained with the
models \textsc{QGSJet01c}, \textsc{QGSJetII.3} and \textsc{Epos~1.61}. 
This study is limited to the case of proton primaries, since
the Glauber model used for iron primaries does not fit into the
framework of any of these models.

We find that the impact of the changes of the extrapolation of
hadronic multiparticle production on air shower predictions is very similar for
all models.
Only \textsc{Epos~1.61} is found to behave somewhat different. The most striking difference between
\textsc{Epos} and the other models is found for RMS($X_{\rm max}$),
see Fig.~\ref{modelDepXmaxRMS}.
All model predictions seem to be virtually independent of the multiplicity, only
\textsc{Epos} shows a clear trend to increased shower fluctuations for
higher multiplicities.
Also the correlation of the elasticity and RMS($X_{\rm max}$) is
stronger in \textsc{Epos} in comparison to the other
models. Furthermore, an increase of the cross section within
\textsc{Epos} does not lead to a significant reduction of the shower
fluctuations in $X_{\rm max}$. This indicates a different contribution
of the residual fluctuations caused by the secondary multiplicity in
\textsc{Epos} than in the other models (c.f.\
Eq.~(\ref{eq:heitlerVar})).

The muon number itself is known to be one of the shower observables
with the largest inherent modelling uncertainties. At the same time it
is an excellent example of how model-independent the relation between
modifications of the
interaction characteristics to air shower observables is, see
Fig.~\ref{modelDepNmuMean}. While the underlying large differences in
the model predictions are clearly visible, the dependence on the
modifications is very similar. The largest differences can be
observed for the elasticity where, similar to the case of RMS($X_{\rm
  max}$), the effect in \textsc{Epos} is larger than for all
other models.



\section{Conclusion and Outlook}
We developed an ad hoc model to modify features of hadronic interaction
in the simulation of extensive air showers. The algorithms
for these modifications were tested using high statistics
data sets of simulated p-air interactions at fixed
energy. We then used this ad hoc model to explore the impact of
different extrapolations of
hadronic interaction characteristics on the predictions for air shower development and
typical air shower observables.

For simplicity it is
assumed that the properties of hadronic interactions are sufficiently
well known at energies below $\unit[10^{15}]{eV}$. Only the
extrapolation of these features to higher energies are changed, while 
ensuring a steady transition of the description of hadronic particle production
from low to high energy.

Our study is mainly based on the interaction model
\textsc{Sibyll~2.1}. This choice is driven by the fact that our model
for introducing modifications to nucleus-air interactions is founded
on the Glauber model and the semi-superposition model as implemented
in \textsc{Sibyll}. The dependence of the results on the particular
interaction model is small. Cross checks with \textsc{QGSJet01c},
\textsc{QGSJetII.3}, and \textsc{Epos~1.61} show that predictions for
air showers change consistently in dependence on the modification we
introduce for hadronic interactions. Some deviations from a universal
behaviour of the models is found for \textsc{Epos}, being most likely
related to the qualitatively different structure of this model.

Many of the observed dependences of air shower predictions on features
of hadronic multiparticle production can be understood qualitatively
within the simple Heitler model of particle cascades and its extension
by Matthews. The most relevant results on
the impact of different extrapolations of hadronic interaction
features on air shower observables are
\begin{itemize}
 \item 
  The longitudinal air shower fluctuations, as described
  by RMS($X_{\rm max}$), depend mainly on the cross section and 
  less strongly on the elasticity.  This makes fluctuations in
  $X_{\rm max}$ a good parameter to study
  hadronic cross sections at ultra-high energies.

\item 
  The electron number is negatively correlated with the multiplicity, 
  whereas the muon number shows a positive correlation.
  The secondary particle multiplicity
  provides a powerful handle to change the electron vs. muon number
  ratio in air showers.

\item 
  The invisible energy fraction is affected by at most $\pm$0.02 by 
  uncertainties of the extrapolation of hadronic interaction features
  to air shower energies.

\item
  To increase the muon number in air showers the
  multiplicity of interactions can be increased, or the charge-ratio
  decreased. A change of the predicted muon number in air showers of
  more than 30\,\% seems almost impossible to obtain by only modifying
  high energy interactions in the simulation as described here.

\end{itemize}

It turns out that the influence of the extrapolation of hadronic
interaction features on air shower observables and the dependence of
these observables on the mass of the primary particle are of similar
magnitude. Therefore it is very difficult to use shower measurements to
estimate the characteristics of hadronic multiparticle production at
ultra-high energy. External knowledge on the primary mass composition
will be needed for reliable estimates. Such information could come
from models of the sources of ultra-high energy cosmic rays or the
observation of magnetic deflection of the particles in the
Galaxy. In addition the suppression of the cosmic ray flux above
$\unit[10^{19.7}]{eV}$ offers a unique window to a mono- or
bi-elemental cosmic ray beam: almost only proton or iron particles are
expected to arrive at Earth~\cite{Harari:2006uy,Allard:2008gj} at
ultra-high energy from distant sources.

The measurement of hadronic multiparticle production at LHC energy has
the potential to significantly reduce the spread of the model
divergences at cosmic ray energies. If the information most relevant
to the air shower development will be extracted from the LHC data,
this will very likely lead to a breakthrough in the analysis of air
shower data.  Already now experiments at the LHC measured multiplicity
pseudorapidity distributions in the central event region very
accurately. However, the really important measurements for the
understanding of air shower cascades will be: leading particle
effects, multiplicities in forward direction and hadronic cross
sections. These measurements are very difficult to perform--mostly because they
require to measure the very forward directed particle production--and
will thus take much more time to be measured to high precision.

\begin{acknowledgments}
  The authors want to thank their colleagues from the Pierre Auger
  Collaboration, and especially Prof.\ P.\ Sommers and Dr.\ T.\ Pierog, 
  for numerous discussions.  The simulations needed for this
  work were performed at the Campus-Grid~\footnote{\texttt{www.campusgrid.de}} at the
  Steinbuch Centre for Computing of the KIT.
  This work has been supported in part by
  Bundesministerium f{\"u}r Bildung und Forschung (BMBF) grant No.\
  05A08VK1. 
\end{acknowledgments}

\vspace*{-.5cm}
\begin{footnotesize}
  \bibliographystyle{article}
  {\raggedright
    \bibliography{article}
  }
\end{footnotesize}

\clearpage

\begin{appendix}

%
\section{Modified Characteristics of Hadronic Interactions}
\label{app:mod}
\begin{figure*}[bth!] 
  \includegraphics[width=.33\linewidth]{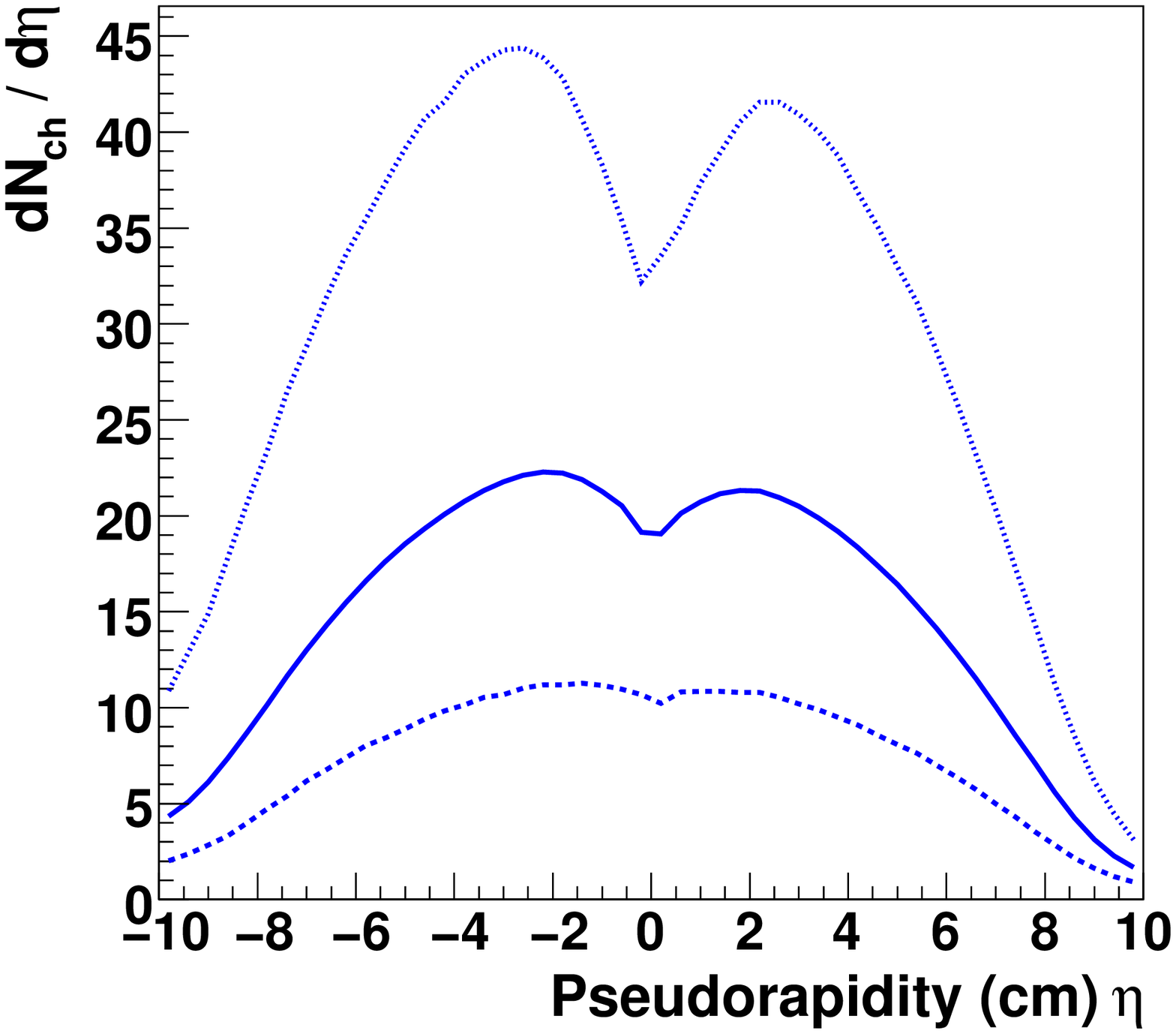}~
  \includegraphics[width=.33\linewidth]{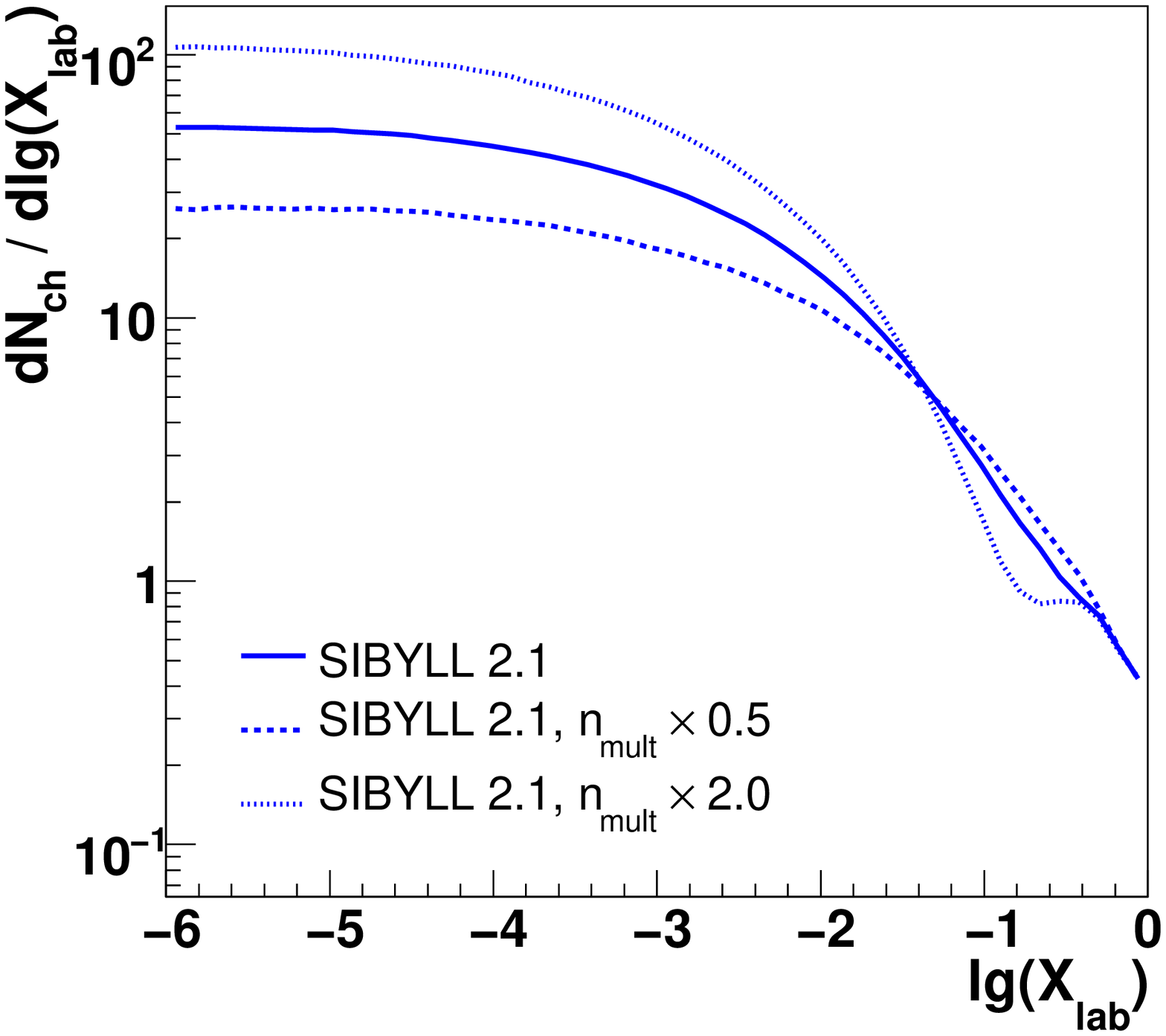}~
  \includegraphics[width=.33\linewidth]{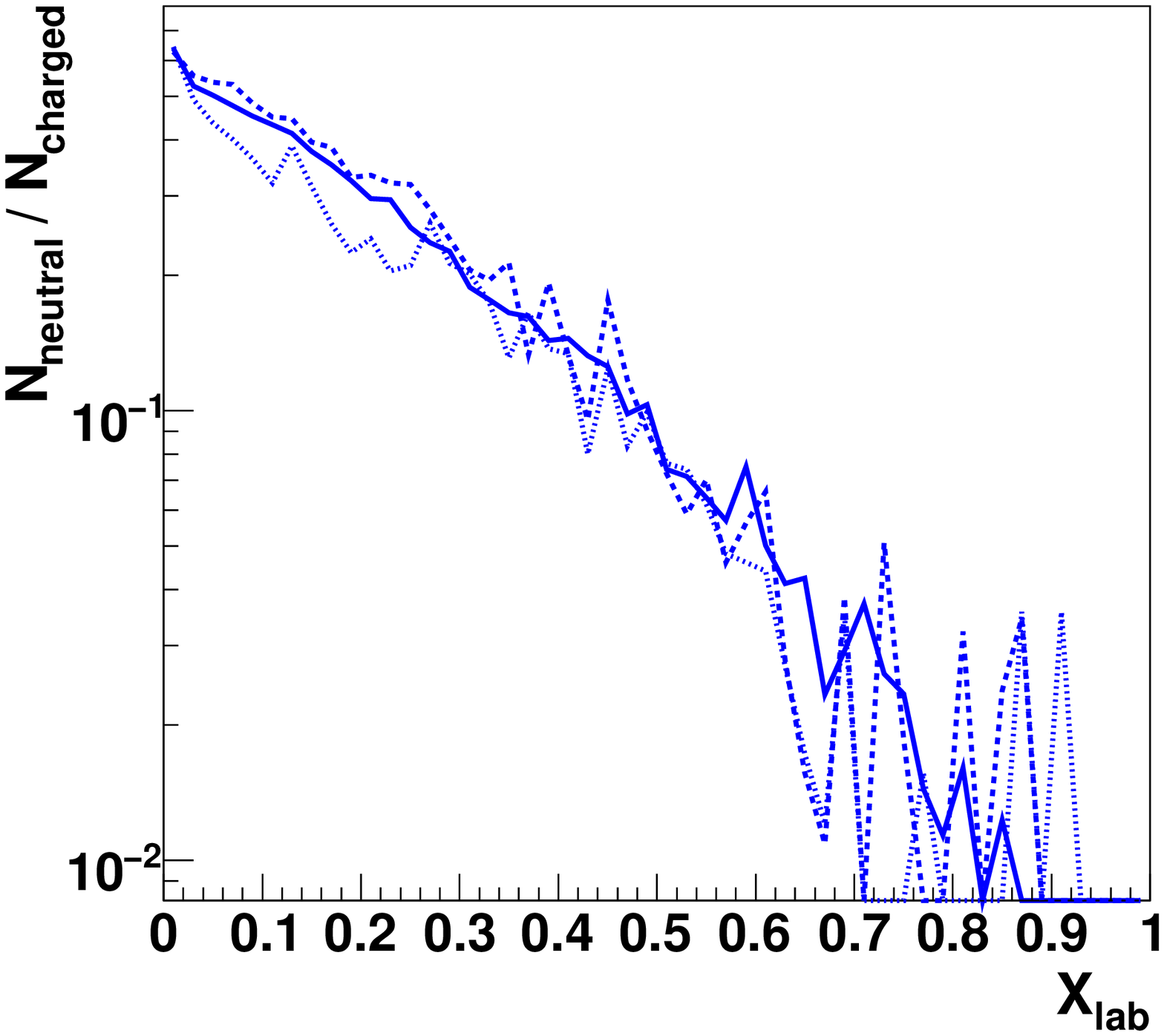}
  \caption{Impact of modifying the multiplicity in proton-air collision
    at $10^{19}\,$eV on the resulting distribution of secondaries.}
  \label{fig:modMult}
\end{figure*} 
\begin{figure*}[tbh] 
  \includegraphics[width=.33\linewidth]{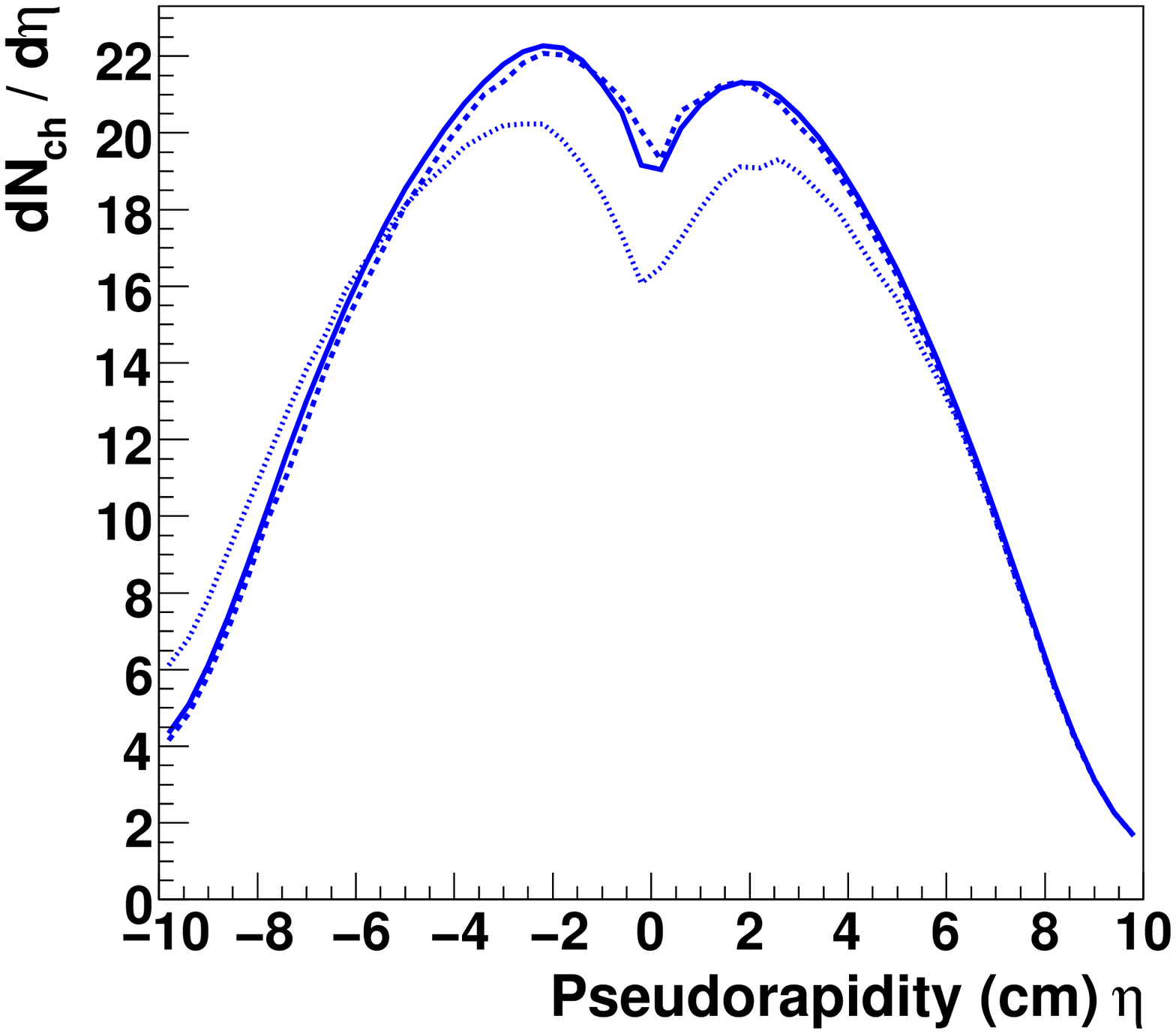}~
  \includegraphics[width=.33\linewidth]{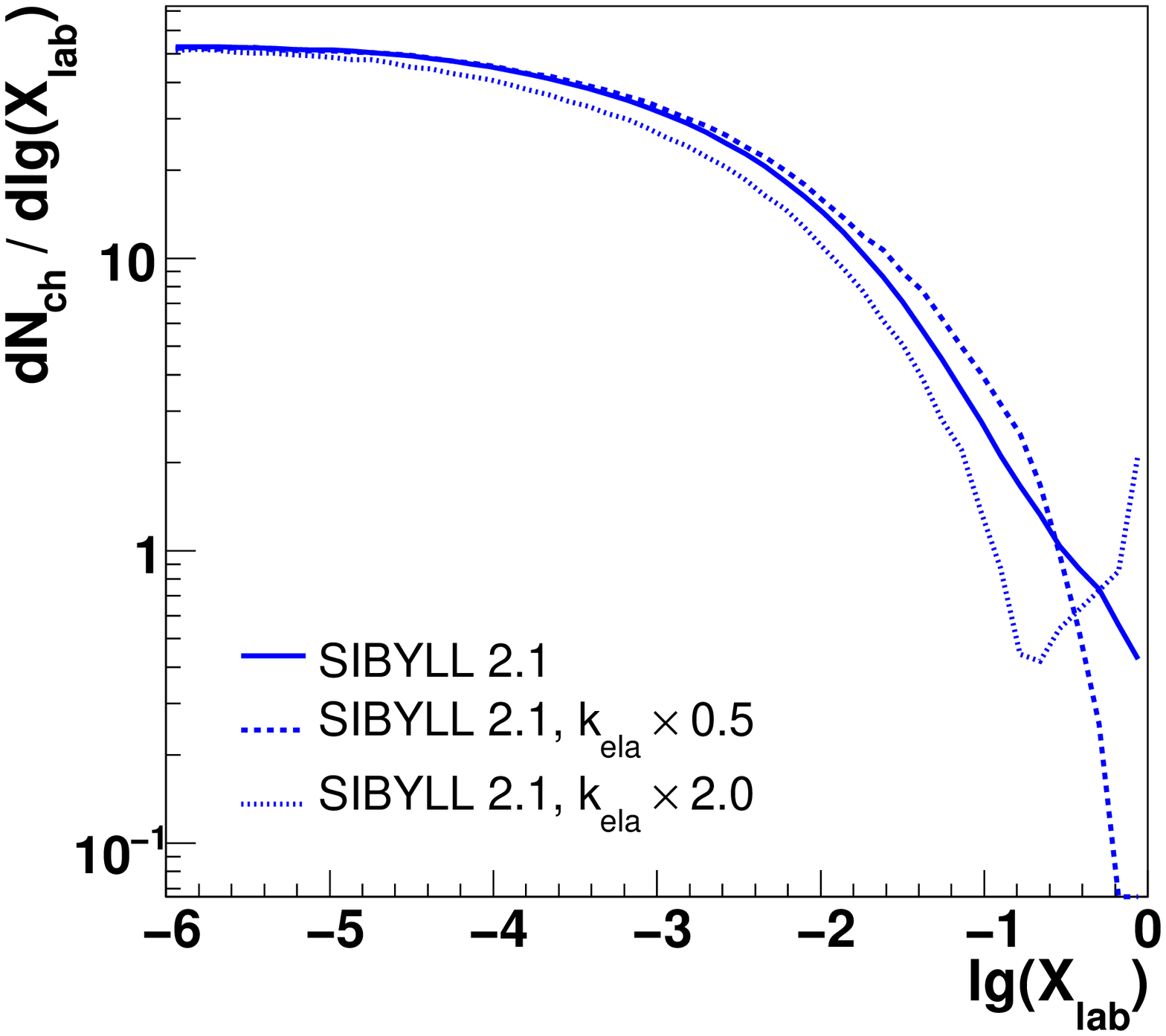}~
  \includegraphics[width=.33\linewidth]{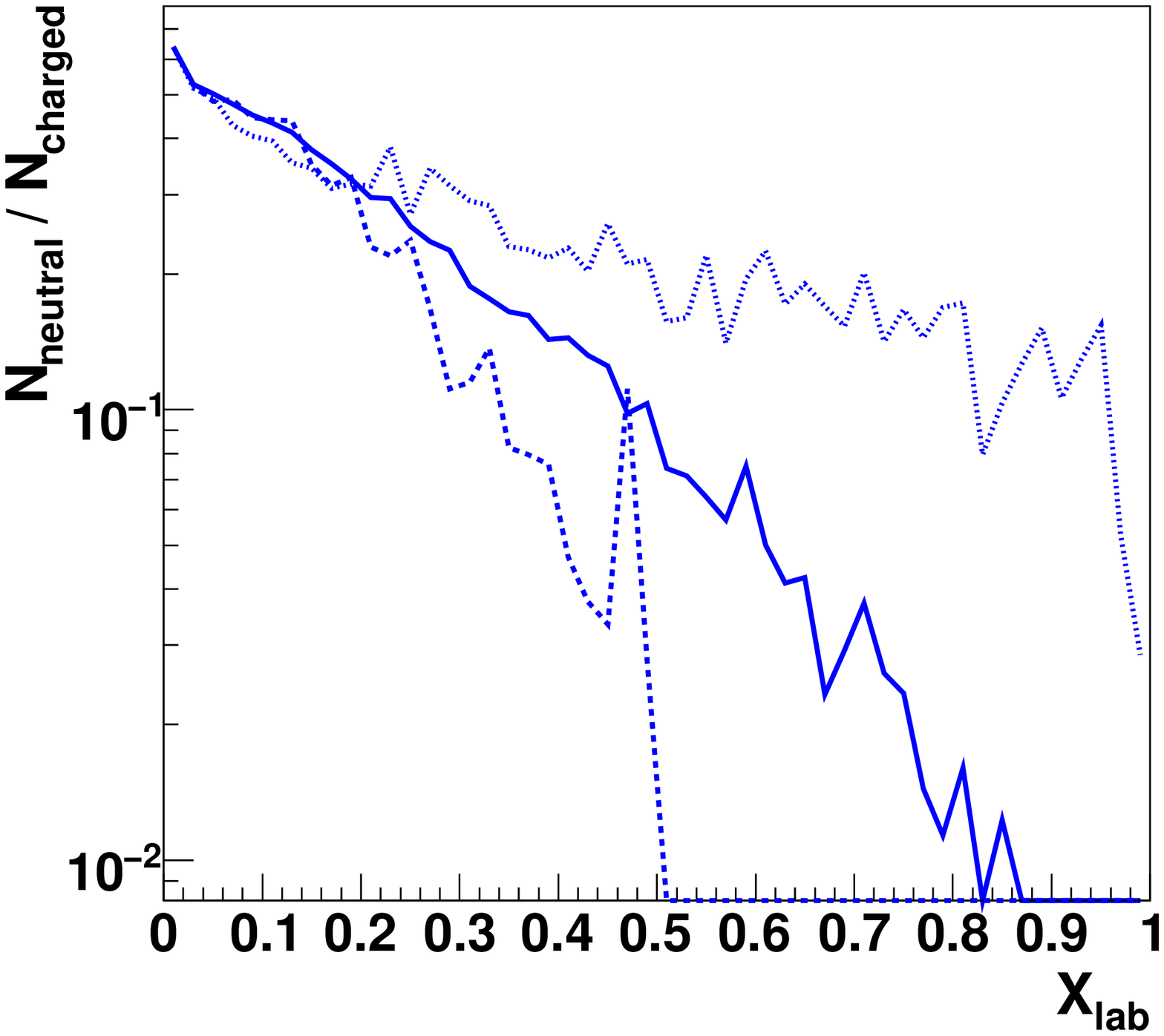}
  \caption{Impact of modifying the elasticity in proton-air collision
    at $10^{19}\,$eV on the resulting distribution of secondaries.}
  \label{fig:modEla}
\end{figure*} 
\begin{figure*}[t]
  \includegraphics[width=.33\linewidth]{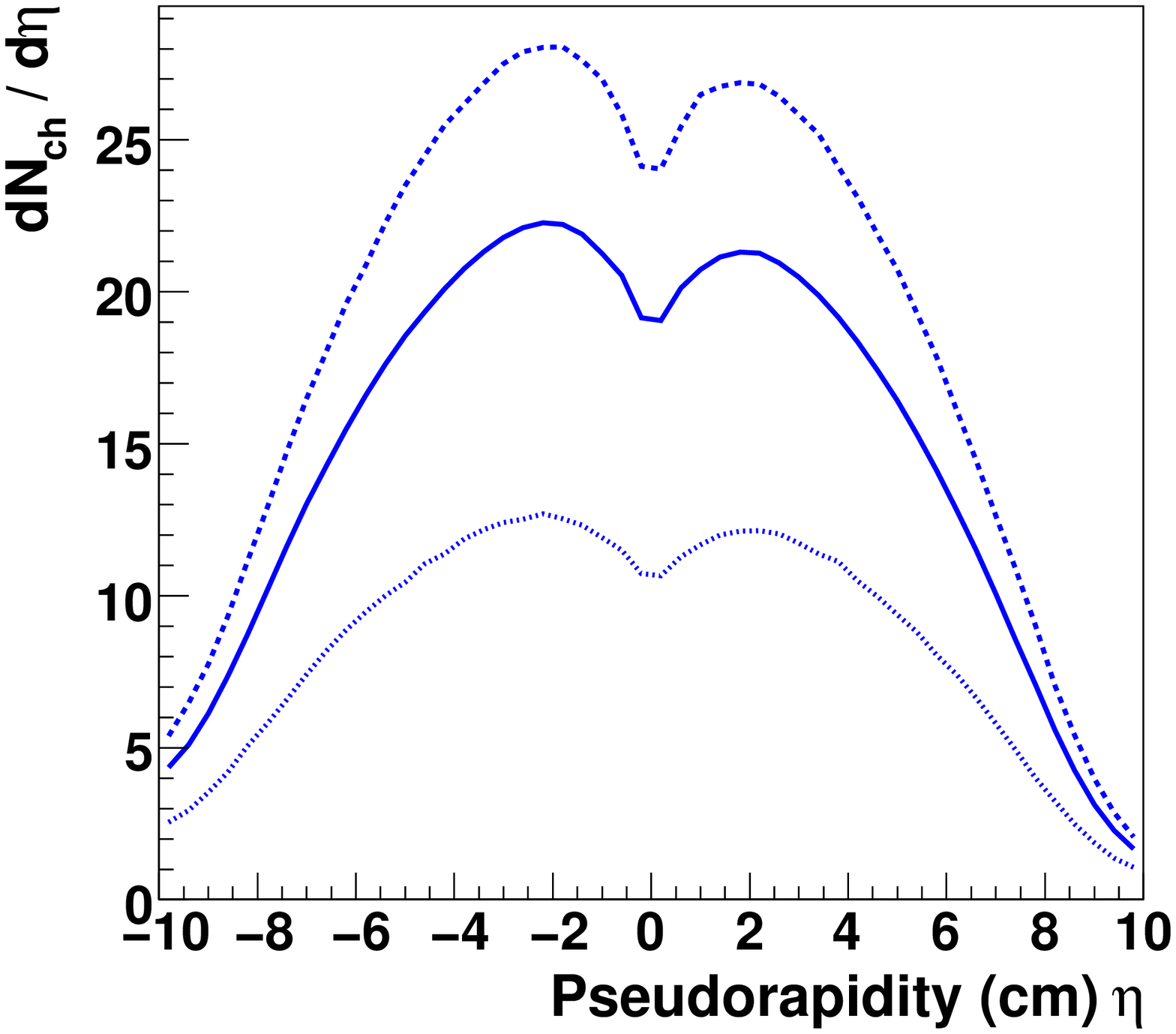}~
  \includegraphics[width=.33\linewidth]{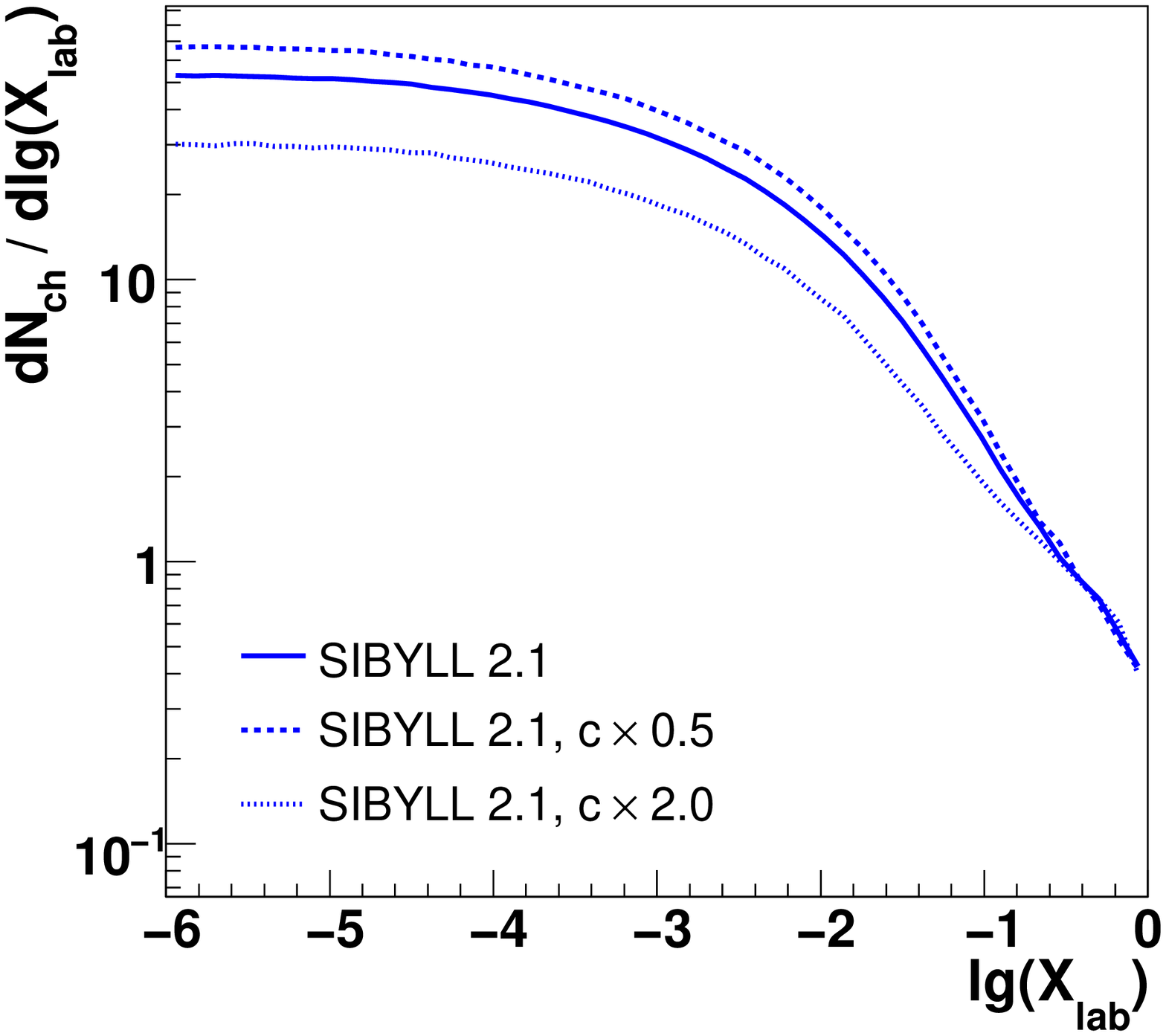}~
  \includegraphics[width=.33\linewidth]{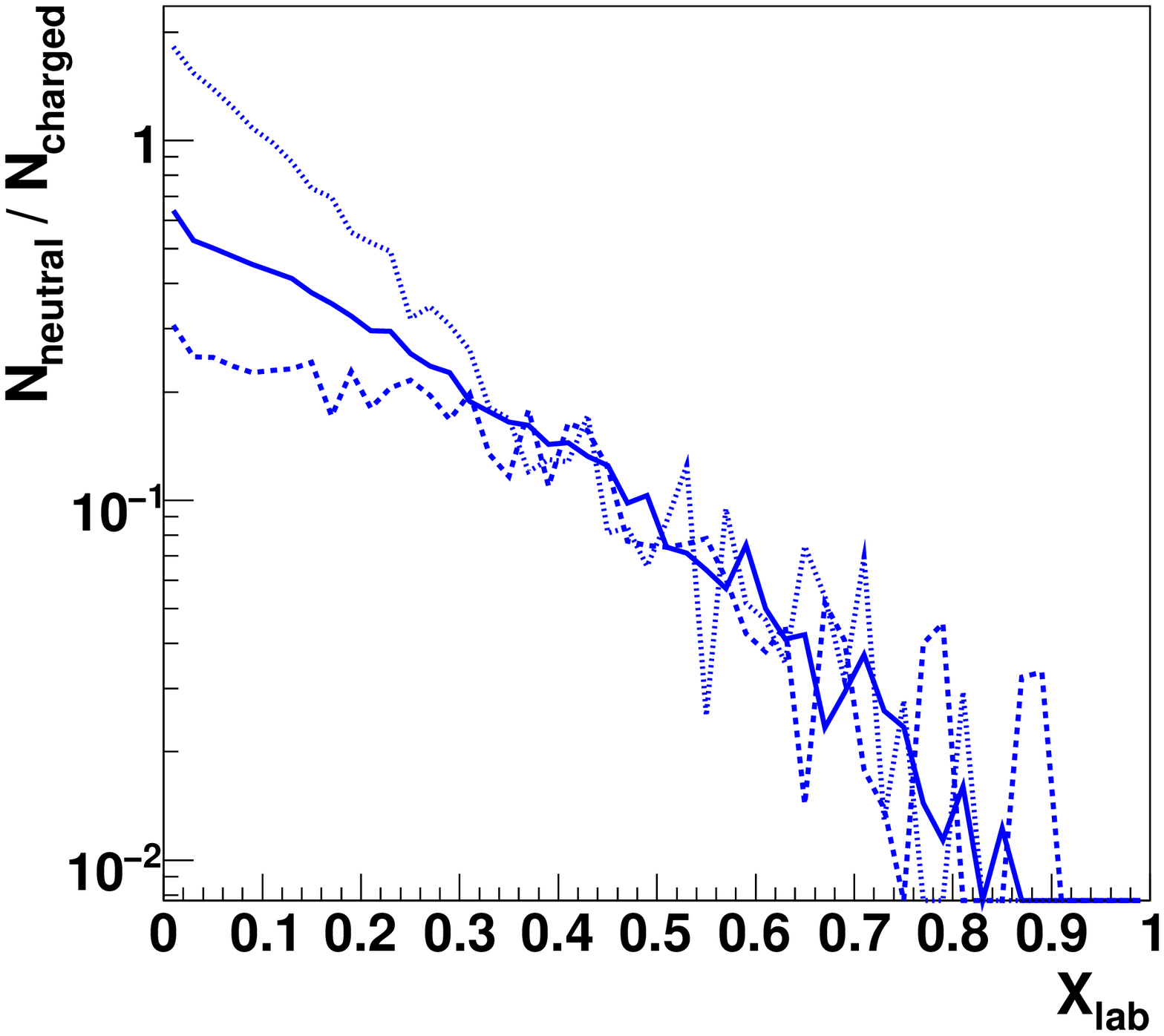}
  \caption{Impact of modifying the charge ratio in proton-air collision
    at $10^{19}\,$eV on the resulting distribution of secondaries.}
  \label{fig:modChargeRatio}
\end{figure*}
The CONEX air shower simulation program was modified allowing us to
change specific properties of the particle production characteristics
of the underlying high energy event generators.

The resampling algorithms are written in order to specifically change
individual properties of the secondary particle distributions, while
conserving all of the important features like total energy, charge,
particle types, energy fractions in different particle types and the
leading particle as far as possible.  Since CONEX is a 1D-EAS
simulation code, no special efforts are made to conserve $p_\perp$. If
the energy $E$ of a secondary particle is altered to yield $E_{\rm
  mod}$ its momentum $\vec{p}$ is re-computed as $\vec{p}_{\rm
  mod}=\vec{p}\,\sqrt{E_{\rm mod}^2-m^2}/|\vec{p}|$.

The algorithms are applied to the secondary particles of interactions in
the laboratory frame of the air shower. This is the
frame that is most relevant for the air shower development, and i.e.\
conservation of the total energy is paramount within this frame.

In the following we will describe in detail how the algorithms work to
achieve the modification of specific interaction features.
All algorithms are controlled by the energy dependent modification
factor $f(E,\,f_{19})$, see Eqs.~(\ref{eqn:modifier}) and
(\ref{eqn:MODIFIER}).

For the discussion we use the quantities pseudorapidity
$\eta=-\ln\tan\theta/2$, $X_{\rm lab}=E/E_0$ and the number of charged
secondaries, $N_{\rm ch}$.

%
\subsection{Cross Sections}
Modifying the cross section can be implemented straightforwardly.  No
secondary particle resampling is needed. Only the extrapolated value
of all hadronic cross sections has to be multiplied by $f(E,\,f_{19})$
\begin{equation}
  \label{fig:app:cxmod}
  \sigma^{\rm mod}=\sigma^{\rm orgi}\;f(E,\,f_{19})\;.
\end{equation}
See Fig.~\ref{f:SigmaModifiedCrossSection} for an example of this
modification.  The rescaling is not only done for the primary particle,
and hence $\sigma_{\rm p-air}$, but for all corresponding hadronic
interaction cross sections in the EAS above the chosen transition
energy of $\unit[10^{15}]{eV}$.

%
\subsection{Secondary Multiplicity}
To change the multiplicity of secondaries, first of all, the leading
particle needs to be excluded to prevent a change of the elasticity at
the same time. The remaining particles are grouped together with
respect to their type: nucleons, pions, kaons, photons, electrons and
muons.  To obtain a different multiplicity, particles are removed or
duplicated at random from these groups depending on the aim of
increasing ($f(E,\,f_{19})>1$) or decreasing ($f(E,\,f_{19})<1$)
the particle multiplicity. After the particle number has been adapted to be
\begin{equation}
  n_{\rm mult}^{\rm mod}=n_{\rm mult}^{\rm orig}\; f(E,\,f_{19})
\end{equation}
the \emph{kinetic} energy of all particles is scaled to yield exactly
the same \emph{total} energy as prior to the resampling. Since the
leading particle is spared from all these operations, the elasticity
of the interaction is conserved.  Furthermore, this assures energy
conservation as well as a constant energy ratio between the particle
type groups. To conserve the charge, particles are changed to their
anti-partners within particle type groups, until the total charge
balance is restored up to a maximum charge offset of $\pm1$. Finally,
all particle momenta are re-computed to account for their modified
energies.

This minimalistic approach is optimized to conserve as many
features of the original interaction model as possible. 
The developed method allows us to resample secondary particles of
hadronic interactions in order to increase or decrease the
multiplicity with as few assumptions as possible. No particle energy
spectra or any other modelling is involved. Of course artificial
fluctuations can be introduced since only existing particles are
deleted or duplicated. But the qualitative shape of the particle
energy spectra, the energy ratio between particle groups, the charge
balance and the total energy as well as the leading particle are
preserved as far as possible.

We studied the impact of the multiplicity modification on high
statistics simulations of proton-air collisions at $10^{19}\,$eV. In
Fig.~\ref{fig:modMult}~(left) it can be seen that modifications of the
multiplicity basically just scale the ${\rm d}N_{\rm ch}/{\rm
  d}\eta$-distribution. The fact that the leading particles are
preserved can be clearly seen in Fig.~\ref{fig:modMult}~(middle) close
to $X_{\rm lab}=1$. To compensate this, there occurs an
under-respectively over-shoot of the ${\rm d}N_{\rm ch}/{\rm
  d}X_{\rm lab}$-distribution around $X_{\rm lab}\sim0.1$. At smaller
$X_{\rm lab}$ the scaling of the multiplicity can again be
identified. Modifying the multiplicity has no impact on the resulting
distribution of $N_{\rm neutral}/N_{\rm charged}$.

%
\subsection{Elasticity of Interactions}
To modify the elasticity $\kappa_{\rm el} = 1-k_{\rm inel}=E_{\rm
  leading}/E_{\rm tot}$ to obtain
\begin{equation}
  \kappa_{\rm el}^{\rm mod} = \kappa_{\rm el}^{\rm orig}\,f(E,\,f_{19})
\end{equation}
it is only needed to re-distribute energy between the leading particle
and the rest of the secondaries. Two facts are limiting the range of
possible modifications of the elasticity:
\begin{description}
 \item[lower bound] If the total available energy of all secondaries
   is distributed equally to all secondaries, then the minimal
   achievable elasticity, $\kappa_{\rm el}^{\rm mod}\ge1/n_{\rm mult}$, of
   this ensemble of particles is reached. This limit corresponds to a
   complete vanishing of any leading particle effects. All secondaries
   are emitted with an equal share of the total energy.
  \item[upper bound] Since no secondary particles should get deleted
   the energy that is bound in the \emph{mass} of all secondary
   particles is not accessible to further increase the elasticity of
   the leading particle. So the maximal elasticity is reached when the
   \emph{kinetic} energy of all particles is transferred to the
   leading particle: $\kappa_{\rm el}^{\rm mod}\le\sum\limits_{i}^{\rm
     n_{\rm mult}} E_i^{\rm kin} / E_{\rm tot}$. In the extreme case
   this corresponds to the production of particles with no kinetic
   energy. The leading particle takes the full energy of the
   projectile minus the energy that went into the mass of the other
   secondaries. Since these secondaries carry no kinetic energy they
   are thus not of any relevance for the development of the air shower
   cascade.
\end{description}
By design the total energy $E_{\rm tot}$ is not changed during the
procedure and since no particles are produced in addition to the
existing ones or removed also the
charge balance as well the particle type statistics are not
altered. Because of the changing energy of all involved particles the
particle momenta need to get re-computed at the end.

We studied the impact of modifying the elasticity on a high statistics
simulation of proton-air collision at $10^{19}\,$eV. In
Fig.~\ref{fig:modEla} (left+middle) it is shown that a smaller
elasticity just removes the leading particles close to $X_{\rm lab}=1$
and slightly increases the number of particles at smaller $X_{\rm
  lab}$. In the central region of the collision there is basically no
change. On the other hand, if the elasticity is increased, this leads
to significantly reduced overall particle production, since more energy
is carried away by the leading particles. This shows also clearly up
in the central region of the collision. 
The seemingly large effect shown in Fig.~\ref{fig:modEla}~(right) is 
due in part to using a linear $X_{\rm lab}$ axis compared to the logarithmic
axis in Fig.~\ref{fig:modEla}~(middle).

%
\subsection{Charge-Ratio of Pions}
By far most of the hadronic particle production ends up in secondary
pions. Since there are three types of pions, $\pi^0$,
$\pi^+$ and $\pi^-$, roughly a third of the produced particles are of
each of the pion types. The \textit{charge ratio} defined as
\begin{equation}
  c=\frac{n_{\pi^0}}{n_{\pi^0}+n_{\pi^+}+n_{\pi^-}}\,,
\end{equation}
with $n_{\pi^{\rm x}}$ being the number of pions of type $\pi^{\rm x}$,
and is of the order of 1/3. Since neutral pions decay almost
instantly producing $2\gamma$, this is the main channel to transfer
energy into the electromagnetic cascade of the shower. The charged
pions, on the other side, carry on the hadronic shower cascade (shower
core) and eventually decay into muons.

We modify the charge ratio to finally yield
\begin{equation}
  c^{\rm mod} = c^{\rm orig}\,f(E,\,f_{19})\;,
\end{equation}
by switching neutral into charged pions ($f(E,\,f_{19})<1$) or vice versa
($f(E,\,f_{19})>1$). We apply the following scheme:
\begin{description}
\item[positive projectiles] $\pi^-$ $\leftrightarrow$ $\pi^0$
\item[negative projectiles] $\pi^+$ $\leftrightarrow$ $\pi^0$
\item[neutral projectiles]  $\pi^\pm$ $\leftrightarrow$ $\pi^0$.
\end{description}
Furthermore, in all cases the \textit{leading} particle is excluded
and thus fully preserved.

Due to this logic there are limitations in the modification of
the charge ratio of the form
\begin{equation}
  0 < c^{\rm mod} < \frac{n_{\pi^0}+n_{\pi^{\rm src}}}{n_{\pi^0}+n_{\pi^-}+n_{\pi^+}}\,,
\end{equation}
where $n_{\pi^{\rm src}}$ is, depending on the projectile, $n_{\pi^+}$ for
negative, $n_{\pi^-}$ for positive, and $n_{\pi^+}+n_{\pi^-}$ for
neutral projectiles.
Given this scheme we find the probability for
\begin{description}
\item [$f(E,\,f_{19})>1$] to switch one $\pi^0 \rightarrow \pi^{\rm src}$ is:\\
  $p=(n_{\pi^0}-f(E,\,f_{19})\,n_{\pi^0})/n_{\pi^0}=1-f(E,\,f_{19})$
\end{description}
and
\begin{description}
\item [$f(E,\,f_{19})<1$] to switch one $\pi^{\rm src} \rightarrow \pi^0$ is:\\
  $p=(f(E,\,f_{19})\,n_{\pi^0}-\,n_{\pi^0})/n_{\pi^{\rm src}}$\;.
\end{description}

Since changing the pion type of individual particles also slightly
changes their rest mass, all momenta need to be recomputated given the
new rest mass.

We studied the impact of modifying the charge ratio on a high
statistics simulation of proton-air collision at $10^{19}\,$eV. In
Fig.~\ref{fig:modChargeRatio} (left) it is shown that the charge ratio
just scales the ${\rm d}N_{\rm ch}/{\rm d}\eta$-distribution, while in
Fig.~\ref{fig:modChargeRatio}~(middle) it is also clear that our
strategy to preserve the leading particle effect works, since there is
no impact on ${\rm d}N_{\rm ch}/{\rm d}X_{\rm lab}$ close to $X_{\rm
  lab}=1$. Similarly Fig.~\ref{fig:modChargeRatio}~(right) illustrates
how the leading particles at high $X_{\rm lab}$ are preserved and the
charge ratio only changes at small $X_{\rm lab}$.

%
\section{Nucleus Projectiles}
\label{app:modnuc}
\begin{figure}[b]
  \centering
  \includegraphics[width=\linewidth]{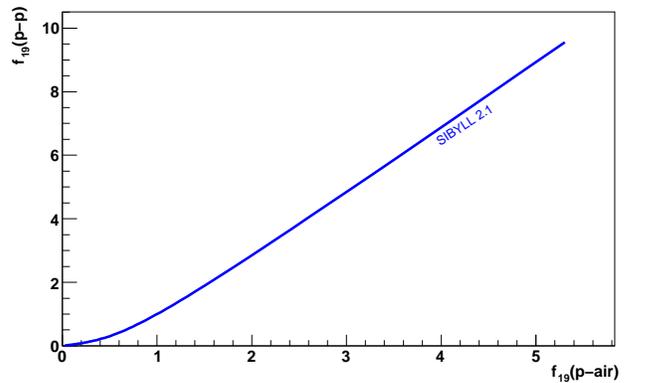}
  \caption{Conversion from $f_{19}$ to $f_{19}^{\rm p-p}$ using the
    Glauber theory within \textsc{Sibyll}.}
  \label{fig:app:f19}
\end{figure}
For projectiles that are nuclei of $A$ nucleons the
modification of interaction characteristics are performed within the
Glauber theory~\cite{Glauber:1955qq,Glauber:1970jm}. With the
semi-superposition model~\cite{Engel:1992vf} it is calculated how many
of the nucleons of the projectile nucleus, $N_{\rm wounded}$, are
taking part in the interaction with the target nucleus (air). The
interaction of these nucleons with the target are then computed
individually, and can thus be rescaled by the same techniques as
introduced in Appendix~\ref{app:mod}, at the reduced projectile energy
of $E_0/A$.

The cross sections for nucleus primaries are also computed within the
Glauber framework. The nucleus-nucleus cross section
is computed based on the fundamental parameters $\sigma_{\rm
  tot}^{\rm p-p}$, $\sigma_{\rm el}^{\rm p-p}$ and $B_{\rm el}$, which are
scaled for our purpose to
\begin{align}
  \sigma_{\rm tot\,,mod}^{\rm p-p} &= f(E,\, f_{19}^{\rm p-p}) \; \sigma_{\rm tot}^{\rm p-p}\\
  \sigma_{\rm ela,\,mod}^{\rm p-p} &= f(E, \,f_{19}^{\rm p-p}) \; \sigma_{\rm el}^{\rm p-p}\\
   B_{\rm ela,\,mod} &= f(E,\, f_{19}^{\rm p-p}) \; B_{\rm el} \;, 
\end{align}
where $f_{19}^{\rm p-p}$ is chosen to yield the following equation
\begin{equation}
  f_{19} = \frac{\sigma^{\rm p-air}_{\rm prod}(\sigma_{\rm tot,\,mod}^{\rm p-p},\,\sigma_{\rm ela,\,mod}^{\rm p-p}\,,B_{\rm ela,\,mod})}{\sigma^{\rm p-air}_{\rm prod}(\sigma_{\rm tot}^{\rm p-p},\,\sigma_{\rm el}^{\rm p-p}\,,B_{\rm el})} \;.
\end{equation}
The proton-air cross sections are all computed in the Glauber
framework as implemented in \textsc{Sibyll}. This approach assures
consistency with the rest of the hadronic cross sections in the air
shower that are modified according to Eq.~(\ref{fig:app:cxmod}).  The
relation between $f_{19}$ and the rescaling parameter of the
fundamental proton-proton interaction parameters $f_{19}^{\rm p-p}$ is
shown in Fig.~\ref{fig:app:f19}.

\end{appendix}

\end{document}